\documentclass[%
 reprint,
 twocolumn,
superscriptaddress,
nofootinbib,
 amsmath,amssymb,
 prd,
]{revtex4-2}

\usepackage{graphicx}% Include figure files
\usepackage{dcolumn}% Align table columns on decimal point
\usepackage{bm}% bold math
\usepackage{color}
\usepackage{siunitx}
\usepackage[colorlinks=true,citecolor=blue,urlcolor=blue]{hyperref}
\usepackage[dvipsnames]{xcolor}
\usepackage{lineno}
\usepackage{bbm}
\usepackage{orcidlink}
\usepackage{url}
\usepackage{ulem} %Remove me after editing
\usepackage{acronym}
\usepackage{float}

\newcommand{\pbar}{\bar{\cal P}}
\newcommand{\hf}[2]{\bar{H}_{#1}^{#2}(f; \boldsymbol{\theta^\prime}_{#1})}
\newcommand{\hfinj}[2]{\bar{H}_{#1}^{#2}(f; \{\vec{\theta}^\mathrm{inj}\})}

\newcommand{\lmax}{\ell_\text{max}}

\newcommand{\orf}[2]{\gamma_{#1}^{#2}(f,t)}
\newcommand{\fisher}[2]{\Gamma_{#1}^{#2}}
\newcommand{\LnOSig}{\ln\mathcal{O}^\mathrm{SIG}_\mathrm{N}}
\newcommand{\LnOAniso}{\ln\mathcal{O}^{\ell=7}_{\ell=0}}
\newcommand{\LnONgr}{\ln\mathcal{O}^\mathrm{NGR}_\mathrm{GR}}

\newcommand*{\diff}{\,\mathrm{d}}
\newcommand*{\eexp}{\mathrm{e}}

\newcommand{\refref}[1]{Ref.~\cite{#1}}
\newcommand{\referenceref}[1]{Reference~\cite{#1}}
\newcommand{\refsref}[1]{Refs.~\cite{#1}}
\newcommand{\referencesref}[1]{References~\cite{#1}}

\newcommand{\tableref}[1]{Table~\ref{#1}}
\renewcommand{\eqref}[1]{Eq.~(\ref{#1})}

\newcommand{\eqsref}[2]{Eqs.~(\ref{#1}) and (\ref{#2})}

\newcommand{\figref}[1]{Fig.~\ref{#1}}
\newcommand{\figsref}[2]{Figs.~\ref{#1} and \ref{#2}}
\newcommand{\figureref}[1]{Figure~\ref{#1}}
\newcommand{\figuresref}[2]{Figures~\ref{#1} and \ref{#2}}
\newcommand{\secref}[1]{Sec.~\ref{#1}}
\newcommand{\sectionref}[1]{Section~\ref{#1}}
\newcommand{\appref}[1]{Appendix~\ref{#1}}

\usepackage[acronym]{glossaries}
\makeglossaries
\begin{document}

\preprint{APS/123-QED}

\title{Extension of the Bayesian searches
for anisotropic stochastic\\gravitational-wave background with nontensorial polarizations}% Force line breaks with \\

\author{Leo Tsukada}
\email{leo.tsukada@ligo.org}
\affiliation{Department of Physics, The Pennsylvania State University, University Park, Pennsylvania 16802, USA}
\affiliation{Institute for Gravitation and the Cosmos, The Pennsylvania State University, University Park, Pennsylvania 16802, USA}
\date{\today}% It is always \today, today,

\begin{abstract}
    The recent announcement of strong evidence for a stochastic
    gravitational-wave background (SGWB) by various pulsar timing array
    collaborations has highlighted this signal as a promising candidate for
    future observations. Despite its nondetection by ground-based detectors
    such as Advanced LIGO and Advanced Virgo, \citet{tom_nongr_method} developed a Bayesian formalism to search for an
    isotropic SGWB with nontensorial polarizations, imposing constraints on
    signal amplitude in those components that violate general relativity using
    LIGO's data. Since our ultimate aim is to estimate the spatial distribution
    of gravitational-wave sources, we have extended this existing method to
    allow for anisotropic components in signal models. We then examined the
    potential benefits from including these additional components. Using
    injection campaigns, we found that introducing anisotropic components into a
    signal model led to more significant identification of the signal itself and
    violations of general relativity. Moreover, the results of our Bayesian
    parameter estimation suggested that anisotropic components aid in breaking
    down degeneracies between different polarization components, allowing us to
    infer model parameters more precisely than through an isotropic analysis. In
    contrast, constraints on signal amplitude remained comparable in the absence
    of such a signal. Although these results might depend on the assumed source
    distribution on the sky, such as the Galactic plane, the formalism presented
    in this work has laid a foundation for establishing a generalized Bayesian
    analysis for an SGWB, including its anisotropies and nontensorial
    polarizations.
\end{abstract}

\maketitle

%% Acronyms
\acrodef{gw}[GW]{gravitational wave}
\acrodef{bh}[BH]{black hole}
\acrodef{cbc}[CBC]{compact binary coalescence}
\acrodef{gr}[GR]{general relativity}
\acrodef{bbh}[BBH]{binary black hole}
\acrodef{bns}[BNS]{binary neutron star}
\acrodef{imbh}[IMBH]{intermediate-mass black hole}
\acrodef{smbh}[SMBH]{super-massive black hole}
\acrodef{ligo}[LIGO]{the Laser Interferometer Gravitational-wave Observatory}
\acrodef{lvk}[LVK]{the LIGO Scientific, Virgo and KAGRA Collaboration}
\acrodef{o2}[O2]{the second observing run}
\acrodef{o3}[O3]{the third observing run}
\acrodef{o3a}[O3a]{the first half of the third observing run}
\acrodef{o4}[O4]{the fourth observing run}
\acrodef{o5}[O5]{the fifth observing run}
\acrodef{snr}[SNR]{signal-to-noise ratio}
\acrodef{csd}[CSD]{cross spectral density}
\acrodef{psd}[PSD]{power spectral density}
\acrodef{pdf}[PDF]{probability density function}
\acrodef{gwtc}[GWTC]{Gravitational Wave Transient Catalog}
\acrodef{msp}[MSP]{milli-second pulsar}
\acrodef{sgwb}[SGWB]{stochastic gravitational-wave background}
\acrodef{bpl}[BPL]{broken power law}
\acrodef{pl}[PL]{power law}
\acrodef{sph}[SPH]{spherical harmonics}
\acrodef{orf}[ORF]{overlap reduction function}
\acrodef{pta}[PTA]{pulsar timing array}
%\tableofcontents

\section{Introduction}
In recent years, there has been a dramatic expansion in the field of \ac{gw}
astronomy, driven by the consistent detection of such signals by
\ac{lvk}~\cite{gwtc-1,gwtc-2,gwtc-2.1, gwtc-3}. The \acp{gw} observed to date
have come from \ac{cbc} such as
\ac{bbh}~\cite{gw150914,gw151226,gw170104,gw170608,gw170814,gw190412,gw190521,
gw190814} and \ac{bns}~\cite{gw170817, gw190425}. Detailed studies of these
astrophysical events have enhanced our understanding of general relativity,
nuclear physics, and the astrophysical processes involved in such mergers, thus
providing a unique perspective into the Universe.  Moreover, the recent
announcement regarding strong evidence for an \ac{sgwb} by several \ac{pta}
collaborations~\cite{Agazie2023-iq,Reardon2023-pj,Xu2023-yo,Antoniadis2023-zr}
has propelled this field forward with the potential detection of an \ac{sgwb},
introducing a potential new avenue in the nHz frequency band to observe
\acp{gw}.

An \ac{sgwb} is the incoherent superposition of \acp{gw} emitted from numerous
sources, which are too faint to be individually resolved (see
e.g.,~\cite{arianna_review} for a detailed review).  It is primarily composed of
astrophysical sources such as \ac{smbh}
binaries~\cite{Rajagopal_1995,Jaffe_2003,Wyithe_2003,Sesana_2004,McWilliams_2014,Burke-Spolaor:2019aa},
targeted by the \ac{pta} collaborations, along with stellar \acp{bbh} or
\acp{bns} \cite{tania,
Regimbau_2022,Banagiri:2020kqd,Payne:2020pmc,Stiskalek:2020wbj}, and supernovae
\cite{2009MNRAS.398..293M,2010MNRAS.409L.132Z,PhysRevD.72.084001,PhysRevD.73.104024,marocmodi}.
Alternatively, cosmological sources, including signals emitted during the
inflationary
era~\cite{1994PhRvD..50.1157B,1979JETPL..30..682S,2007PhRvL..99v1301E,2012PhRvD..85b3525B,2012PhRvD..85b3534C,Lopez:2013mqa,1997PhRvD..55..435T,2006JCAP...04..010E,axion_inflation},
phase transitions in the early
Universe~\cite{vonHarling:2019gme,Dev:2016feu,Marzola:2017jzl}, and primordial
\acp{bh} \cite{MandicEA_2016,SasakiEA_2016,Wang:2016ana,Miller:2020kmv}, can
contribute to the \ac{sgwb}. Despite the promising results from the PTA
collaborations, ground-based detectors like \ac{ligo}~\cite{ligo} and
Virgo~\cite{virgo} have not made any substantial detections. This is largely
because they are tuned to higher frequency GWs produced by events within a
relatively nearby Universe, whereas a considerable portion of the \ac{sgwb}
signal falls into the low-frequency regime. Since May 2023, with further
improved sensitivities of \ac{ligo} and Virgo, the expanded detector network
including KAGRA~\cite{kagra} has begun the fourth observing run, aiming for the
novel discovery of a \ac{sgwb} signal.

As one of the physical implications from \ac{gw} observations, there have been
extensive studies that examine the theories of gravity by investigating the
violation of general relativity in \ac{gw}
observations~\cite{gw150914_tgr,gw170817_tgr,gwtc-1_tgr, gwtc-2_tgr,
gwtc-3_tgr}. These include (but are not limited to) the parametrized tests of
the post-Newtonian
coefficients~\cite{Blanchet1994-ux,Blanchet1995-eh,Arun2006-cl,Arun2006-yn,Yunes2009-mu},
the speed of \acp{gw}~\cite{gw170817_grb}, black-hole nature of the merger
remnant~\cite{Healy2017-ml,Hofmann_undated-qu,Jimenez-Forteza2017-ai,Westerweck2018-iq,Uchikata2019-rw,Abedi2017-je,ashton2016comments}
and, most relevant to this work, the nontensorial \ac{gw}
polarizations~\cite{isi2017probing,gw190814,gw170817_tgr,Takeda2021-ro,Takeda2022-lb,Chatziioannou2012-aj,Hagihara2019-il,Pang2020-lz}.
Regarding the \ac{gw} polarization analyses, \refref{isi2017probing}
demonstrates the method to search for a \ac{gw} signal from a \ac{cbc}, which is
purely polarized with either one of three polarizations, i.e. \textit{scalar} or
\textit{vector} or \textit{tensor}. This method was adopted for the analysis of
GW170814~\cite{gw170814} and GW170817~\cite{gw170817_tgr}, which were
consequently reanalyzed with modified waveforms~\cite{Takeda2021-ro} and
tensor-scalar mixed polarization model~\cite{Takeda2022-lb}, respectively.
Alternatively, \refref{Chatziioannou2012-aj} demonstrates the use of
\textit{null streams}, constructed by a \ac{gw} model with the tensor
polarizations, as a model-independent way to assess the \ac{gw} polarization
contents violating the prediction from general relativity. They find that the
independent measurement of each \ac{gw} polarization for transient \ac{gw}
signals requires the same number of detectors as the polarization modes to
measure, which leaves up to five detectors\footnote{Note that the scalar
longitudinal and breathing modes indicate complete degeneracy in their antenna
responses (see \eqref{eq:antenna_longi_breath}) of the current ground-based
detectors.}. \referencesref{Hagihara2019-il,Pang2020-lz} apply this approach to
GW170817 and obtain a constraint of \ac{gw} amplitude for the vector
polarization or p-values for the nontensorial polarization hypothesis.

As opposed to the limitation for polarization measurements in such a transient
\ac{gw} signal, unpolarized \acp{sgwb}
\footnote{\refref{Isi:2018aa} pointed out
that one cannot necessarily apply the model-independent formalism as some of the
assumptions break down in beyond-GR theories, e.g. Chern-Simons gravity.
Nevertheless, this is beyond the scope of this paper and we leave this as a
potential avenue to pursue in the future.}
allow for measuring or constraining
the background amplitude for each of the three \ac{gw} polarizations separately.
In the context of ground-based \ac{gw} detectors, \refref{Nishizawa:2009aa}
first explores the detectability of an \ac{sgwb} with nontensorial
polarizations based on a frequentist approach by deriving \ac{snr} for arbitrary
polarization contents, finding that the separate measurement of generic \ac{gw}
polarization contents require at least three detectors. Alternatively,
\refref{tom_nongr_method} demonstrates a \textit{Bayesian} framework to search
for each polarization component of an isotropic \ac{sgwb} with a broadband
frequency spectrum. More specifically, it provides a statistical prescription to
assess the presence of nontensorial polarizations based on a Bayes factor as
well as posterior results of given model parameters as a way of parameter
estimation. Following this method, \refsref{nongr_sgwb_lvk, o2_iso, o3_iso}
search for nontensorial polarization components of an \ac{sgwb} in \ac{lvk}'s
dataset up to the third observing run.

In this work, we extend the Bayesian framework introduced by
\refref{tom_nongr_method} to account for \textit{anisotropies} of an \ac{sgwb}.
Several searches for anisotropic \acp{sgwb} contributed from either pointlike
or extended sources have been developed and performed for \ac{lvk}'s dataset.
While pixel-wise radiometer
methods~\cite{radio_method,Ballmer2006LIGOIO,Mitra_2008_radio_method2} are
intended for pointlike sources, one of the common approaches for extended
sources is the use of the \ac{sph} expansion of an anisotropic \ac{sgwb} signal
in \ac{csd} $C(f,t)$, which reads~\cite{sph_methods, sph_pe}
\begin{align}
    \label{eq:csd_mean}
    \langle C(f, t)\rangle=\sum_{\mu\in\{(\ell, m)\}}\gamma_\mu(f, t) \mathcal{P}_{\mu}(f).
\end{align}
Here, $\orf{\mu}{}$ is referred to as the \ac{orf}~\cite{ORF} projected onto the
\ac{sph} basis represented by $\mu=(\ell, m)$~\cite{allen-ottewill}, and
$\mathcal{P}_{\mu}(f)$ is a generic form of a spectral model with the
anisotropic distribution.  The Greek subscript implies the summation across the
\ac{sph} modes.  \referenceref{sph_pe} incorporates this expression into the Bayesian
framework of an isotropic \ac{sgwb} to include its anisotropies and provide
detection statistics as well as posterior results for an assumed source
distribution on the sky. \referenceref{Chung2023-eb} explores the possibility to reconstruct a \ac{sgwb} intensity map using marginalized posterior distributions. In this paper, we essentially combine the
formalisms developed by \refsref{tom_nongr_method,sph_pe} and establish a
generalized Bayesian analysis for an \ac{sgwb} including anisotropies
\textit{and} nontensorial polarizations.

This paper is structured as follows. \sectionref{sec:orf} presents the role and
derivation of \ac{orf} in the context of ground-based \ac{gw} detectors. In
particular, we compute the \ac{orf} projected onto the \ac{sph} basis for the
nontensorial polarizations as a key ingredient for this analysis.  In
\secref{sec:bayes}, focusing on the Bayesian formalism as a versatile tool for
the data analysis in this work, we provide an overview of the Bayesian framework
and applications to the detection and parameter estimation of an anisotopic
\ac{sgwb}.  Proceeding to \secref{sec:sigid}, we demonstrate how to identify an
anisotropic \ac{sgwb} signal with the nontensorial polarizations using a
simulated noise dataset for the two \ac{ligo} detectors. Specifically, a
synthetic anisotropic \ac{sgwb} signal is injected into the dataset and
recovered with either isotropic or anisotropic \ac{sgwb} model and we compare
the detection capability between the two cases.  Finally, in \secref{sec:pe}, we
investigate the results of parameter estimation to quantify the measurement
accuracy for the model parameters of an injected \ac{sgwb} signal such as
amplitude and frequency spectrum, demonstrating various benefits gained by
incorporating higher \ac{sph} modes into a signal model.

\section{Overlap reduction function}
\label{sec:orf}
\subsection{Formalism}
\label{sec:orf-form}
As mentioned in \eqref{eq:csd_mean}, an \ac{sgwb} signal is imprinted in the
\ac{csd} estimator, multiplied with a scaling factor and phase shift, which are
determined by a baseline's geometry and relative orientation and represented by
the \ac{orf} $\gamma(f)$.  Hence, this encodes the sensitivity of a given
baseline to an \ac{sgwb} as a function of frequencies.  In the context of
searches for anisotropic \acp{sgwb}, in general the \ac{orf} also depends on
time due to the Earth's rotation with respect to the cosmological rest frame.
Specifically, at a particular point on the sky $\boldsymbol{\hat{\Omega}}$, this
is given by\footnote{Note that the expression in \eqref{eq:orf_pixel} is
normalized such that in the case of a colocated and coaligned detector pair
$\int\diff\boldsymbol{\hat{\Omega}}\gamma(f,t,\boldsymbol{\hat{\Omega}})=1$ at
any frequency for the \textit{tensor} polarization, and hence the those for
vector or scalar polarizations represent values relative to the tensor
counterpart.}~\cite{ORF,romano2017detection,Allen:1999aa,pystoch_sph}
\begin{align}
    \label{eq:orf_pixel}
    \gamma^I(f, t, \boldsymbol{\hat{\Omega}})=\frac{5}{8\pi}\sum_A F_{1}^A(\boldsymbol{\hat{\Omega}}, t) F_{2}^A(\boldsymbol{\hat{\Omega}}, t) \eexp^{2 \pi i f \frac{\boldsymbol{\hat{\Omega}} \cdot \boldsymbol{\Delta x}_I(t)}{c}},
\end{align}
where $ \boldsymbol{\Delta x}_I(t)$ is the time-dependent separation vector
between the two detectors in a baseline $I$. 

$F_{i}^A(\boldsymbol{\hat{\Omega}}, t)$ is the $i$ th detector's antenna pattern
response function for the polarization $A$, which reads
\begin{align}
    \label{eq:antenna}
    F_{i}^A(\boldsymbol{\hat{\Omega}}, t) = d^{\mu\nu}_i(t)e^A_{\mu\nu}(\boldsymbol{\hat{\Omega}}).
\end{align}
$d^{\mu\nu}_i(t)$ is the response tensor of $i$ th detector as follows
\begin{align}
    \label{eq:det_resp}
    \boldsymbol{d}_i(t)=\frac{1}{2}\left(\boldsymbol{\hat{X}}(t)\otimes\boldsymbol{\hat{X}}(t)-\boldsymbol{\hat{Y}}(t)\otimes\boldsymbol{\hat{Y}}(t)\right),
\end{align}
where $\boldsymbol{\hat{X}}(t),\boldsymbol{\hat{Y}}(t)$ are the time-dependent unit vectors that define the direction of detector's $x$ or $y$-arm respectively, and $\otimes$ represents a tensor product.
$e^A_{\mu\nu}(\boldsymbol{\hat{\Omega}})$ is the polarization tensor, defined
commonly on the \textit{\ac{gw} frame}, in which $\boldsymbol{\hat{\Omega}}$ is
equivalent to the unit vector pointing at a \ac{gw} source, and
$\boldsymbol{\hat{m}}$, $\boldsymbol{\hat{n}}$ are a pair of orthogonal unit
vectors in the plane perpendicular to $\boldsymbol{\hat{\Omega}}$. Using these
bases, for the two modes $A=\{+, \times\}$ expected from general relativity, the
polarization tensors are expressed as
\begin{align}
    \boldsymbol{e}^+(\boldsymbol{\hat{\Omega}}) &= \boldsymbol{\hat{m}}\otimes\boldsymbol{\hat{m}} - \boldsymbol{\hat{n}}\otimes\boldsymbol{\hat{n}}\\
    \boldsymbol{e}^\times(\boldsymbol{\hat{\Omega}}) &= \boldsymbol{\hat{m}}\otimes\boldsymbol{\hat{n}} + \boldsymbol{\hat{n}}\otimes\boldsymbol{\hat{m}}.
\end{align}
As a reference, once putting these together, using Euler angles $(\theta, \phi,
\psi)$ in the \textit{detector frame}, whose $x$ and $y$ axes point toward detector's arms, the antenna pattern
response functions for the two modes read
\begin{align}
    \label{eq:fplus}
    &\begin{aligned}
    F^+(\theta, \phi, \psi)  =&~\frac{1}{2}(1+\cos^2 \theta) \cos 2 \phi \cos 2\psi\\
    &-\cos\theta\sin 2 \phi \sin 2\psi,
    \end{aligned}\\
    \label{eq:fcross}
    &\begin{aligned}
    F^\times(\theta, \phi, \psi)  =&-\frac{1}{2}(1+\cos^2 \theta) \cos 2 \phi \cos 2\psi\\
    &-\cos\theta\sin 2 \phi \cos 2\psi.
    \end{aligned}
\end{align}

To adapt to an \ac{sgwb} search for extended sources, the \ac{orf} is
projected from the pixel basis to the \ac{sph} basis
\begin{align}
    \label{eq:pix2sph}
    \gamma^I_{\ell m}(f, t)=\int_{S^2} \diff \boldsymbol{\hat{\Omega}} \gamma^I(f, t, \boldsymbol{\hat{\Omega}}) Y_{\ell m}^*(\boldsymbol{\hat{\Omega}}),
\end{align}
where $Y_{\ell m}$ is the $(\ell, m)$ mode of the \ac{sph} function defined on
the cartesian coordinate whose $z$-axis points to the Earth's rotation axis.
This \ac{sph}-based \ac{orf} has been analytically and numerically derived for
the tensor polarizations in the literature~\cite{allen-ottewill,pystoch_sph}.
This illustrates sensitivity to an \ac{sgwb} with the spatial scale
characterized by $(\ell, m)$ at each frequency bin for a given baseline's
geometry and orientation.  For example, as described in \appref{app:orf}, the
\ac{orf} for $\ell=1$ or $2$ the global peak lies at frequencies below
\SI{100}{\hertz}, while for $\ell\geq5$ the peak appears at higher frequencies.
This can be understood by the diffraction limit which each baseline is
associated with due to the separation between the two detectors, i.e. the
spatial resolution of signal components with lower frequencies tends to be
limited by some angular resolution.  In what follows, we explore the extension
of the \ac{sph}-based \ac{orf} to nontensorial \ac{gw} polarizations, i.e.
scalar and vector polarizations.

\subsection{Extension to the nontensorial polarizations}
The \ac{orf}'s dependence on the \ac{gw} polarization boils down to the
polarization tensor, $e^A_{\mu\nu}(\boldsymbol{\hat{\Omega}})$. For the
nontensorial polarization modes, these tensors read\footnote{One should note
that there exist two different conventions of
$\boldsymbol{e}^\ell(\boldsymbol{\hat{\Omega}})$~\cite{tom_nongr_method,Nishizawa:2009aa}, and we adopt that of Callister
\textit{et al.} in \refref{tom_nongr_method} to follow their formalism
consistently.}
\begin{align}
    \label{eq:e_b}
    \boldsymbol{e}^b(\boldsymbol{\hat{\Omega}}) &= \boldsymbol{\hat{m}}\otimes\boldsymbol{\hat{m}} + \boldsymbol{\hat{n}}\otimes\boldsymbol{\hat{n}}\\
    \label{eq:e_l}
    \boldsymbol{e}^\ell(\boldsymbol{\hat{\Omega}}) &= \boldsymbol{\hat{\Omega}}\otimes\boldsymbol{\hat{\Omega}}\\
    \label{eq:e_x}
    \boldsymbol{e}^x(\boldsymbol{\hat{\Omega}}) &= \boldsymbol{\hat{m}}\otimes\boldsymbol{\hat{\Omega}} + \boldsymbol{\hat{\Omega}}\otimes\boldsymbol{\hat{m}}\\
    \label{eq:e_y}
    \boldsymbol{e}^y(\boldsymbol{\hat{\Omega}}) &= \boldsymbol{\hat{n}}\otimes\boldsymbol{\hat{\Omega}} + \boldsymbol{\hat{\Omega}}\otimes\boldsymbol{\hat{n}},
\end{align}
where the superscripts denote breathing ($b$), longitudinal ($\ell$), vector-x
($x$) and vector-y ($y$) modes, respectively. These polarization tensors lead to
the antenna pattern response function for each polarization as follows
\begin{align}
    \label{eq:antenna_longi_breath}
    F^b(\theta, \phi, \psi) & =-F^{\ell}(\theta, \phi, \psi) =\frac{1}{2} \sin ^2 \theta \cos 2 \phi,\\
    F^x(\theta, \phi, \psi) & =\sin \theta(\cos \theta \cos 2 \phi \cos \psi-\sin 2 \phi \sin \psi), \\
    F^y(\theta, \phi, \psi) & =-\sin \theta(\cos \theta \cos 2 \phi \sin \psi+\sin 2 \phi \cos \psi),
\end{align}
in terms of Euler angles on the detector frame, similar to
\eqsref{eq:fplus}{eq:fcross}.
\begin{figure*}[htbp]
    \includegraphics[width=\textwidth]{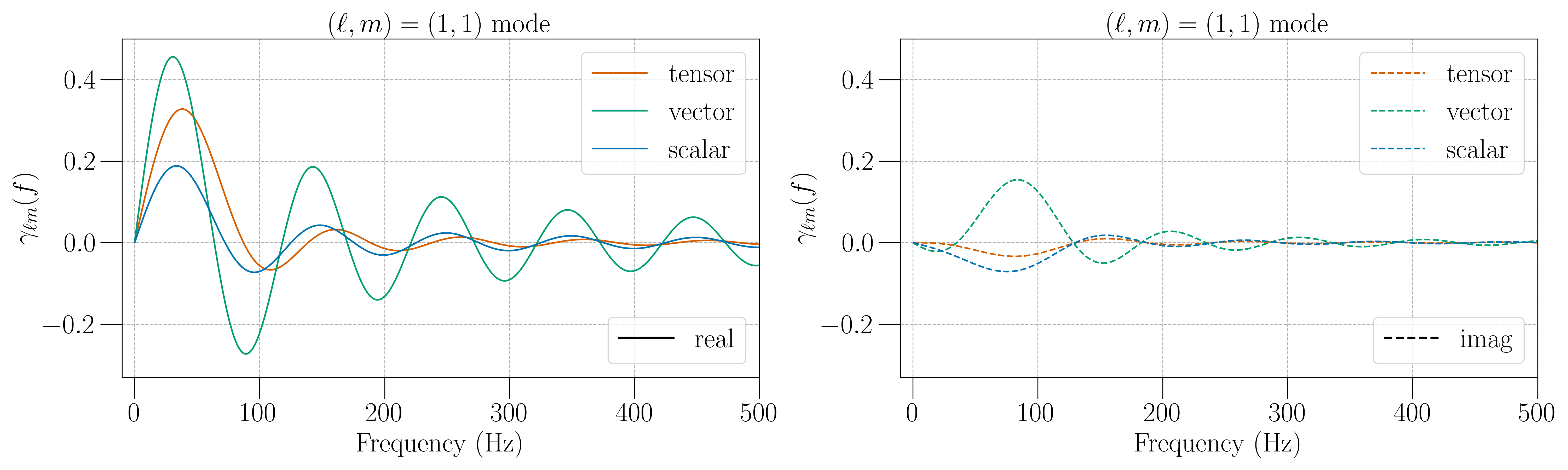}
    \caption{\label{fig:orf_11_hl}
    \ac{sph}-based \ac{orf} of the two \ac{ligo}
    detector pair for each polarization family over \SIrange{0}{500}{\hertz} for
    $(\ell, m)=(1,1)$ mode. The left plot shows the real part of the \ac{orf},
    while the right one shows its imaginary part.}
\end{figure*}
\begin{figure*}[htbp]
    \includegraphics[width=\textwidth]{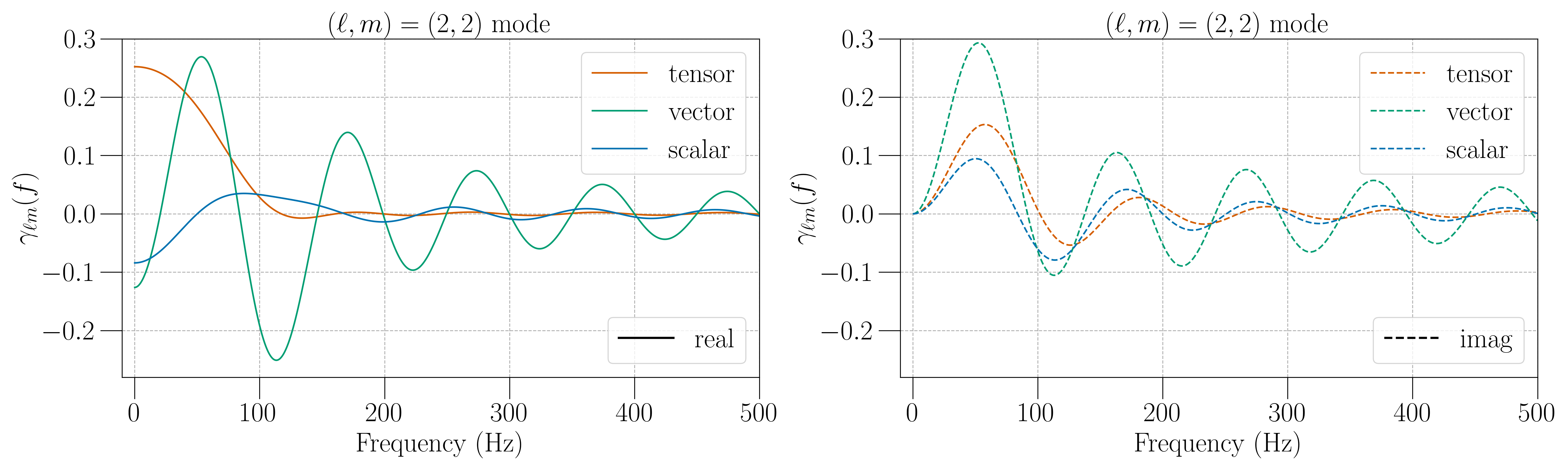}
    \caption{\label{fig:orf_22_hl} \ac{sph}-based \ac{orf} of the two \ac{ligo}
    detector pair for each polarization family over \SIrange{0}{500}{\hertz} for
    $(\ell, m)=(2,2)$ mode. The left plot shows the real part of the \ac{orf},
    while the right one shows its imaginary part.}
\end{figure*}

Since we assume unpolarized \ac{gw} power within each of the vector and scalar
polarizations similarly to the tensor polarizations, the pixel-based \ac{orf}
for the nontensorial polarizations takes the same form as \eqref{eq:orf_pixel}.
Instead of the analytical expression described in \refref{allen-ottewill}, we
follow the numerical approach shown in \refsref{pystoch_sph, pystoch}. First, we
substitute the polarization tensors shown in \eqsref{eq:e_b}{eq:e_l},
\eqsref{eq:e_x}{eq:e_y} into \eqref{eq:antenna} respectively to compute the
pixel-based \ac{orf} with respect to the \texttt{HEALPix} pixelization of
$n_\mathrm{side}=16$ (i.e. $12\times16^2=3072$ pixels) \cite{HEALPix}. Subsequently, following \eqref{eq:pix2sph},
we convert it from the pixel basis to the \ac{sph} basis using the
\texttt{healpy} package \cite{healpy}.

We note that, in the coordinate system where its $z$-axis points toward the
Earth's rotational axis, the time dependence of the \ac{orf} due to the Earth's
rotation is equivalent to the change in the azimuthal angle, and hence 
\begin{align}
    \gamma^I_{\ell m}(f, t)=\gamma_{\ell m}^I(f, t_\mathrm{ref}) \eexp^{i m 2 \pi (t-t_\mathrm{ref}) / T},
\end{align}
where $T$ is the period of the Earth's rotation. This factorization allows us to
compute the \ac{sph}-based \ac{orf} at a reference time $\gamma^I_{\ell m}(f,
t_\mathrm{ref})$ and the time-dependent phase separately.  Eventually,
$\gamma^I_{\ell m}(f, t_\mathrm{ref})$ is then multiplied by the phase factor to
account for the proper time shift and to construct a two-dimensional $f$-$t$ map
, which substantially reduces the computational cost.

We show the \ac{sph}-based \ac{orf} of the two \ac{ligo} detector pair for each
polarization family over \SIrange{0}{500}{\hertz}, e.g. for brevity only $(\ell,
m)=(1,1)$ and $(2,2)$ modes in \figsref{fig:orf_11_hl}{fig:orf_22_hl},
respectively.  Since the \ac{sph}-based \ac{orf} is a complex function in
general, the left plot in either figure shows the real part of each
$\gamma^I_{\ell m}(f, t_\mathrm{ref})$ and the right plot shows its imaginary
part.  Note that $t_\mathrm{ref}$ is adjusted such that the imaginary part of
the tensorial \ac{orf} at $f=\SI{0}{\hertz}$ is zero, which allows for a fair
comparison of the plotted values across the three polarizations at the
consistent reference time.

As mentioned in \secref{sec:orf-form}, at each frequency bin the amplitude of
$\gamma^I_{\ell m}(f, t_\mathrm{ref})$ the sensitivity of the baseline $I$ to
the \ac{sgwb} component whose spatial scale is characterized by the $(\ell, m)$
mode. The frequency range that yields the peaked amplitude is a consequence of
the interplay between the diffraction effect in an observed \ac{sgwb} signal and
the phase canceling between two detectors at higher frequencies. As opposed to
the diffraction limit suppressing sensitivity over lower frequencies, the signal
components at higher frequencies tend to have phases at two detectors
canceled out to some extent when taking crosscorrelation, leading to a decaying
function of the \ac{orf} in general.  This tradeoff is more noticeable for
higher ($\ell, m$) modes, e.g. \appref{app:orf} showing the \ac{orf} for
$(\ell,m)=(10, 10)$ mode, which has the amplitude peaked at around
\SI{200}{\hertz}.

% Also, one can see that, similar to the tensor polarizations, the \acp{orf} for
% the nontensorial polarizations have zeros in both real and imaginary parts.
% This indicate that at the reference time $t_\mathrm{ref}$, those frequency
% components completely cancel out after taking crosscorrelation between the two
% detectors.  These zeros are almost equally spaced in a frequency domain, whose
% interval is disproportional to the light travel time between the two detectors,
% and hence the baseline with larger separation, e.g. Hanford \ac{ligo} and Virgo
% detectors, have more zeros in a given frequency band, as illustrated in
% \appref{app:orf}.

\section{Bayesian Formalism}
\label{sec:bayes}
\subsection{\label{sec:bayes_likelihood} Likelihood}
Following the Bayesian formalism for an anisotropic \ac{sgwb} discussed in
\refref{sph_pe}, here we expand this approach to allow for an inferred signal
model with nontensorial \ac{gw} polarizations or even a mixture of different
polarizations. In general, the mixture model of multiple signal components takes
the form of
\begin{align}
    \sum_i \bar{H}_i(f)\orf{\mu}{A_i} \mathcal{P}^i_\mu,
\end{align}
where for $i$~th signal component $\hf{}{i}$ is the frequency spectrum
normalized at some reference frequency, $A_i$ is its \ac{gw} polarization and
the subscript $\mu$ implies the summation over the \ac{sph} basis
\begin{align}
      \orf{\mu}{A_i}\mathcal{P}^{i}_\mu = \sum_{\ell=0}^{\ell_{\mathrm{max}, i}}\sum^{\ell}_{m=-\ell}\orf{\ell m}{A_i}\mathcal{P}_{\ell m}^{i}.
\end{align}
Note that in general the $\lmax$ value, i.e. spatial cutoff scale in the
\ac{sph} expansion, can be different across signal components.

Following the approach described by \refref{sph_pe}, we normalize the
$\mathcal{P}_{\ell m}$ such that
\begin{align}
    \mathcal{P}_{\ell m} = \epsilon\pbar_{\ell m}\quad
    \textrm{s.t.}\quad\pbar_{00} = \frac{3H_0^2}{2\pi f_\mathrm{ref}^3\sqrt{4{\pi}}}
\end{align}
where $\epsilon$ is an amplitude parameter compatible to $\Omega_\mathrm{ref}$
in the conventional isotropic \ac{sgwb} searches \cite{o1_iso, o2_iso, o3_iso}.
Henceforth, we consider $\pbar_{\ell m}$ to be fixed for each signal component
but still be potentially different across them.  Given the multiple
signal components $\boldsymbol{\mathcal{M}}$, each of which is characterized by
$\bar{H}$, $\gamma_\mu$ and $\pbar_{\mu}$, the Gaussian likelihood reads
\begin{widetext}
    \begin{align}
    \label{eq:likelihood}
    p\left(\left\{C_{ft}\right\} \left|\{\epsilon_i, \boldsymbol{\theta^\prime}_i\} ;\boldsymbol{\mathcal{M}}\right.\right)
        \propto\exp\left\{-\frac{1}{2}\sum_{f, t}\frac{\left|C(f, t)-\sum_i\epsilon_i \hf{i}{} \orf{\mu}{A_i} \pbar^i_\mu\right|^{2}}{P_{1}(f, t) P_{2}(f, t)}
        \right\}.
    \end{align}
\end{widetext}
Here, $P_k(f,t)$ is the \ac{psd} of $k$~th detector, and
$\boldsymbol{\theta^\prime}_i, \epsilon_i$ are a set of model parameters
regarding $\bar{H}(f)$ and the amplitude parameter of the $i$~th signal
component in $\boldsymbol{\mathcal{M}}$, respectively.  $\sum_i$ represents the
summation across $\boldsymbol{\mathcal{M}}$. Note that the case of $\lmax=0$ for
all the signal components reduces to the isotropic search for nontensorial
\ac{gw} polarizations, e.g.  \cite{tom_nongr_method,nongr_sgwb_lvk}. Once a
distribution of likelihood based on \eqref{eq:likelihood} is constructed, it
will be used to produce a posterior probability distribution
$p\left(\{\epsilon_i, \boldsymbol{\theta^\prime}_i\} \left| \left\{C_{ft}\right\};\boldsymbol{\mathcal{M}}\right.\right)$ based on the Bayes' theorem.

\subsection{\label{sec:bayes_odds} Odds ratio}
The Bayesian formalism provides a statistical way to evaluate the preference of
a particular hypothesis over another, so-called \textit{odds ratio} defined as
\begin{align}
    \mathcal{O}_{\mathcal{H}_2}^{\mathcal{H}_1}=\frac{p\left(\mathcal{H}_1 \mid \left\{C_{ft}\right\}\right)}{p\left(\mathcal{H}_2 \mid \left\{C_{ft}\right\}\right)}=\frac{p\left(\left\{C_{ft}\right\} \mid \mathcal{H}_1\right)}{p\left(\left\{C_{ft}\right\} \mid \mathcal{H}_2\right)} \frac{\pi\left(\mathcal{H}_1\right)}{\pi\left(\mathcal{H}_2\right)},
\end{align}
where $\mathcal{H}_1$ is the hypothesis of interest and $\mathcal{H}_2$ is
another one to compare that against. $p\left(\left\{C_{f t}\right\} \mid
\mathcal{H}\right)$ is the Bayesian \textit{evidence} given by
\begin{align}
    p\left(\left\{C_{f t}\right\} \mid \mathcal{H}\right)=\int \mathbf{d}\boldsymbol{\theta} ~p\left(\left\{C_{f t}\right\} \mid\boldsymbol{\theta} ; \mathcal{H}\right) p(\boldsymbol{\theta}),
\end{align}
and $\pi(\boldsymbol{\theta}), \pi(\mathcal{H})$ are the prior probability for
the model parameters $\boldsymbol{\theta}$ and a hypothesis $\mathcal{H}$,
respectively.

In the context of detecting nontensorial \ac{gw} polarizations, similarly to
\refsref{tom_nongr_method,nongr_sgwb_lvk}, we consider the two hypotheses:
\begin{itemize}
  \item GR : only the tensor polarization family is present in the data,
  \item NGR : either scalar or vector polarization family is present in the data.
\end{itemize}
Therefore, denoting $\mathcal{H}_A$ as a subhypothesis of a given combination
(or either) of tensor (T), vector (V) and scalar (S) polarizations present in
data, the GR hypothesis corresponds to $\mathcal{H}_T$, while the NGR hypothesis
consists of $\{\mathcal{H}_\mathrm{S}, \mathcal{H}_\mathrm{V},
\mathcal{H}_\mathrm{TS},$ $ \mathcal{H}_\mathrm{VS}, \mathcal{H}_\mathrm{TV},
\mathcal{H}_\mathrm{TVS}\}$. Accordingly, the odds ratio of the NGR hypothesis
against the GR counterpart reads
\begin{align}
    \label{eq:odds_ngr}
    \mathcal{O}_\mathrm{GR}^\mathrm{NGR}&=\frac{\sum_{A\neq \mathrm{T}}p\left( \left\{C_{ft}\right\} \mid\mathcal{H}_A\right)\pi\left(\mathcal{H}_A\right)}{p\left(\left\{C_{ft}\right\}\mid \mathcal{H}_\mathrm{T} \right)\pi\left(\mathcal{H}_\mathrm{T}\right)}\\
    &=\sum_{A\neq T}\mathcal{O}_{\mathcal{H}_\mathrm{T}}^{\mathcal{H}_A}.
\end{align}
Here, we assign the prior probability equally to each subhypothesis, i.e.
\begin{align}
    \pi\left(\mathcal{H}_\mathrm{A}\right)=\frac{1}{7}\quad ^\forall A\in\{\mathrm{T,S,V,TS,VS,TV,TVS}\}.
\end{align}

\subsection{\label{sec:bayes_implementation} Implementation}
It is computationally challenging to evaluate the likelihood naively based on
\eqref{eq:likelihood} as it involves the integration over frequency and time as
well as summation across the \ac{sph} modes across multiple signal components.
\referenceref{sph_pe} provides an implementation to bypass part of the calculation by precomputing
the time integration. Here, we follow this approach and account for extra
complication due to the presence of multiple signal components.
Specifically, after expanding the exponent of \eqref{eq:likelihood}, the logarithmic likelihood contains the following terms
\begin{align}
    \label{eq:rec_terms}
    \sum_i\epsilon_i\mathrm{Re}[\left(\pbar^i_{\mu}\right)^{*} X^i_{\mu}]
    - \frac{1}{2}\sum_{i,j}\epsilon_i\epsilon_j\left(\pbar^i_{\mu}\right)^{*} \fisher{\mu\nu}{ij} \pbar^j_{\nu}.
\end{align}
Here, the dirty map $X_{\mu}$ is an observable given by \ac{csd} convolved with
detectors' response function and spectrum model $H(f)$, and the Fisher matrix
$\fisher{\mu\nu}{}$ is the covariance matrix of the dirty map. See its full derivation in \refref{sph_methods}.

We note that these quantities are now specific to each signal component as
follows\footnote{The dependency on $f,t$ and $\boldsymbol{\theta^\prime}$ is
omitted for brevity.}
\begin{align}
    \label{eq:dirty_map}
    X^i_{\mu} &=\sum_{f}\bar{H}^i\underbrace{\sum_{t} \frac{\tau\Delta f\left(\gamma^{A_i}_\mu\right)^{*} C_{ft}}{P_{1} P_{2}}}_{\mathrm{precomputed\  for\ }A_i},\\
    \label{eq:fisher_matrix}
    \fisher{\mu\nu}{ij} &=\sum_{f}\bar{H}^i\bar{H}^j\underbrace{\sum_{t} \frac{\tau\Delta f\left(\gamma^{A_i}_\mu\right)^{*} \gamma^{A_j}_\nu}{P_{1} P_{2}}}_{\mathrm{precomputed\ for\ }(A_i, A_j)},
\end{align}
where $\tau$ is the time interval over which the Fourier transform is applied to a time series strain data and $\Delta f$ is the frequency resolution of \ac{csd}. Therefore, the precomputed part in the dirty map (the Fisher
matrix) needs to be stored for each (every combination) of the three \ac{gw}
polarizations, respectively. Subsequently, during the likelihood evaluation the
precomputed part with the associated polarization(s) of a given signal component
or a pair of those is retrieved and used to compute
\eqsref{eq:dirty_map}{eq:fisher_matrix}. This implementation allows the analysis
to explore arbitrary signal-model space with minimal computational cost.

In order to efficiently construct a posterior \ac{pdf} over a multidimensional
parameter space, the pipeline stochastically samples a set of model parameters
defined in a given signal model.  Every time a new sample is drawn, the dirty
map and the Fisher matrix are constructed based on
\eqsref{eq:dirty_map}{eq:fisher_matrix} and eventually the likelihood shown in
\eqref{eq:likelihood}, is evaluated.  This sampling process is repeated until
the Bayesian evidence is computed with sufficient precision. Specifically,
we adopt a nested-sampling algorithm \texttt{Dynesty}~\cite{dynesty},
implemented in the \texttt{Bilby} package.~\cite{bilby1, bilby2}

\section{\label{sec:sigid} signal identification}
\referencesref{sph_pe,tom_nongr_method} show, as described in
\secref{sec:bayes_odds}, that one can use the odds ratio to evaluate the
statistical significance for a signal model of interest. Here, we study the
capability of detecting a signal among the noise and, more importantly, identifying
a signal from the NGR hypothesis as opposed to the GR counterpart.  To this end,
in order to follow the case studies performed in \refref{tom_nongr_method}, in
this section and hereafter we restrict ourselves to considering an anisotropic
\ac{sgwb} injection with only scalar or tensor polarization family, or its
mixture, which is also motivated by scalar-tensor theories~\cite{scalar-tensor}.
Yet, this can be extended to include vector polarizations and the results are
not expected to change drastically.

\subsection{\label{sec:sigid-singlepol} Single polarization}
\begin{figure*}[htbp]
    \centering
    \begin{minipage}[t]{\columnwidth}
        \centering
        \includegraphics[width=\textwidth]{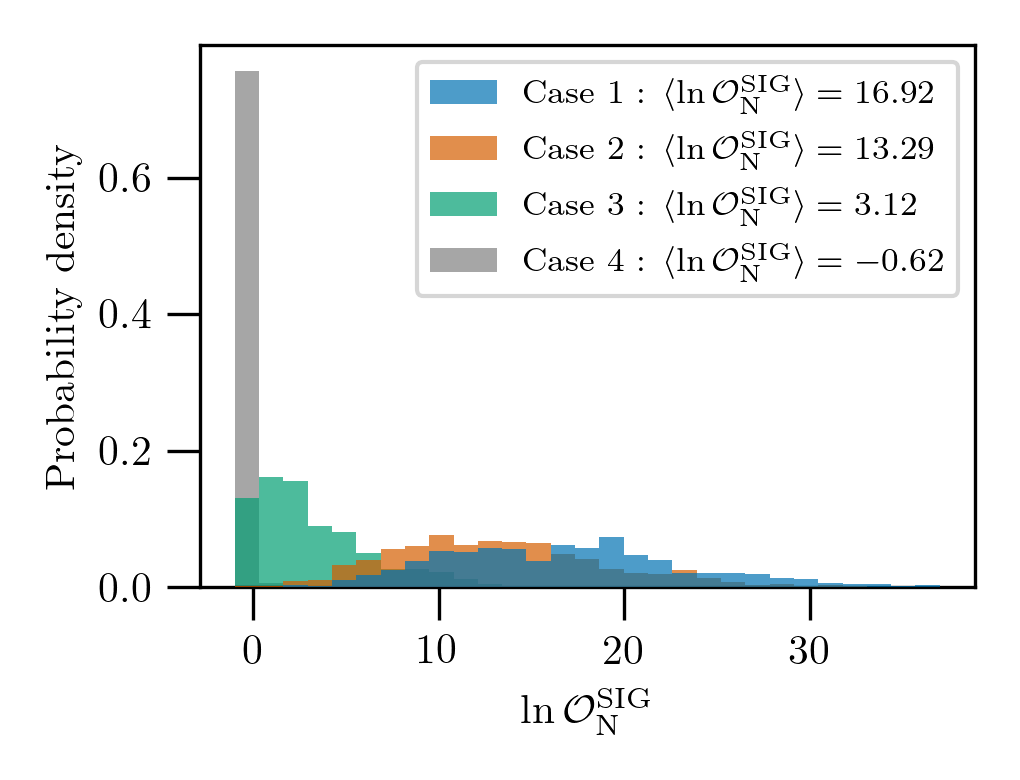}
    \end{minipage}
    \hspace*{\fill}   % <-- added
    \centering
    \begin{minipage}[t]{\columnwidth}
        \centering
        \includegraphics[width=\textwidth]{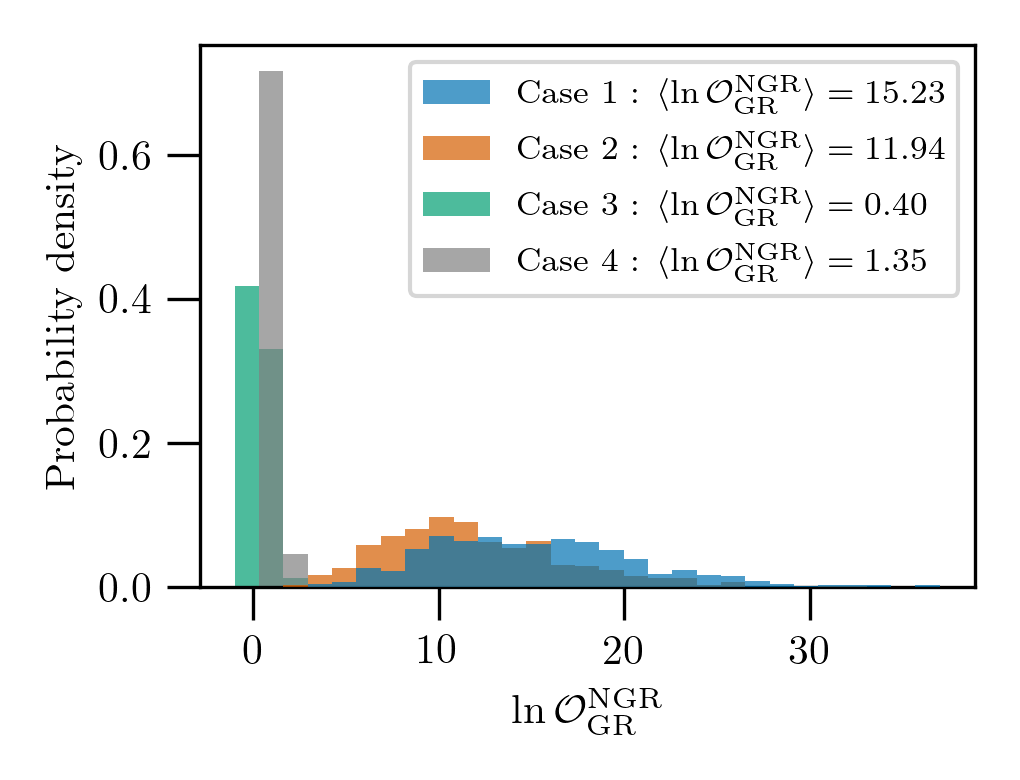}
    \end{minipage}
    \caption{\label{fig:lnO_hist}
    Left : \ac{pdf} of $\LnOSig{}$ between signal and noise
    hypotheses for each of the four cases summarized in
    \tableref{tab:run_config}. Right : \ac{pdf} of $\LnONgr{}$
    between the NGR and GR hypotheses for each of the four cases summarized in
    \tableref{tab:run_config}}
\end{figure*}
We first perform an injection campaign using synthetic \ac{sgwb} signals with
either scalar or tensor polarization. Specifically, the $\pbar_{\ell m}$
distribution is generated from the mock Galactic plane shown in Fig.~1 of
\refref{sph_pe} with $\lmax=5$ and the power-law spectrum $\hf{}{}$ with the index
$\alpha=2/3$. The amplitude factor is determined such that the isotropic
component ($\lmax=0$) of the signal can be recovered with the optimal
\ac{snr}=5, i.e.  $\epsilon\approx\SI{1.8e-8}{}(\SI{4.3e-9}{})$ for the scalar
(tensor) polarization, respectively\footnote{See Eq.~(9) in
\refref{tom_nongr_method} for the definition of the optimal \ac{snr}.}. Our
pipeline injects this signal into simulated noise \ac{csd} between the two
\ac{ligo} detectors of a one-year observation with the design
sensitivity~\cite{observingScene}, and recovers it with the scalar polarization
and $\lmax=0$ or 5.  In addition to these injection runs, we also analyze
simulated data without injections, recovered with scalar polarization and
$\lmax=5$.  \tableref{tab:run_config} summarizes four cases of the configuration. Regarding the
prior \ac{pdf}, a log-uniform distribution and a Gaussian distribution with zero
mean and a standard deviation of 3.5 are used for $\epsilon$ and $\alpha$,
respectively. 
\begin{table}[htbp]
    \centering
{\renewcommand\arraystretch{1.2}
    \begin{tabular}{|c|c|c|c|c|}\hline
      & \multicolumn{2}{c|}{polarization} & \multicolumn{2}{c|}{$\lmax$} \\ \cline{2-5}
       & injection & recovery & injection & recovery\\ \hline\hline
      Case 1 & scalar & scalar & 5 & 5 \\ \hline
      Case 2 & scalar & scalar & 5 & 0 \\ \hline
      Case 3 & tensor & scalar & 5 & 5 \\ \hline
      Case 4 & N/A & scalar & N/A & 5 \\ \hline
    \end{tabular}
}
    \caption{Configurations of the injection campaign}
    \label{tab:run_config}
  \end{table}

An analysis for each case is repeated for 500 different realizations of the
noise \ac{csd} and produces a distribution of the log odds ratio $\LnOSig{}$.
Note that throughout this work the prior odds for both SIG and N models are
assumed to be equal, and hence the odds ratio reduces to the Bayes factor,
\begin{equation}
    \mathcal{B}^{\mathrm{SIG}}_{\mathrm{N}}=\frac{p(\left\{C_{f t}\right\}|\mathrm{SIG})}{p(\left\{C_{f t}\right\}|\mathrm{N})}.
\end{equation}
The left plot in \figref{fig:lnO_hist} shows a \ac{pdf} of $\LnOSig{}$ for each of the four
cases represented by the different colors. One can see that the \ac{pdf} for
\textit{Case 4} (the noninjection run) is narrowly distributed around zero,
while those with the injections extend up to around
$\ln\mathcal{O}^\mathrm{SIG}_\mathrm{N}=30$. Even among these injection runs,
the \acp{pdf} are largely distinct. In particular, the \ac{pdf} of \textit{Case
3} (with the injection of the tensor polarization) is strongly skewed toward
zero and their mean is decreased from \textit{Case 1} (recovered with the scalar
polarization and ) by more than 10 times, which suggests that the signal
recovery using an inconsistent polarization reduces the capability of detecting
the injected signal. Furthermore, apart from the composition of \ac{gw}
polarizations, the choice of $\lmax=0$ (\textit{Case 2}) or 5 (\textit{Case 1})
makes a noticeable impact on the $\LnOSig{}$ \ac{pdf}. This is consistent with
the results demonstrated in \refref{sph_pe} despite the different \ac{gw}
polarization used for each study. More quantitatively speaking, the difference
in the mean values of around 3.7 is in great agreement with the
$\epsilon=10^{-8}$ case of Fig.~8 therein. In summary, one can infer the $\lmax$
value as well as the \ac{gw} polarization by comparing the $\LnOSig{}$ computed
by multiple signal models with the different choices of those hyperparameters.
See a more thorough study in \secref{sec:sigid-mixpol}.

Additionally, we analyze each realization of a dataset using the recovery signal
model, which involves every possible polarization combination and is otherwise
consistent with each of the four configurations listed in
\tableref{tab:run_config}. This allows us to calculate the odds ratio of the NGR
hypothesis against the GR based on \eqref{eq:odds_ngr}.
\figureref{fig:lnO_hist} indicates that overall the \acp{pdf} of $\LnONgr{}$
follows a similar behavior to those of $\LnOSig{}$ except for \textit{Case 3}.
Therefore, the choice of $\lmax$ has a reasonable effect on not only signal
detection but also a statistical test to assess the NGR hypothesis. Regarding
the behavior of \textit{Case 3}, despite the presence of the tensor polarization
alone, the $\LnONgr{}$ does not strongly extend to a negative regime and its
mean value is $\langle\LnONgr{}\rangle=0.40$. This is because no detection of
the nontensorial polarizations only places the upper limit of their \ac{sgwb}
amplitude and does \textit{not} entirely prefer the GR hypothesis, being
consistent with the statement in \refref{tom_nongr_method}.
\subsection{\label{sec:sigid-mixpol} Mixed polarizations}
\begin{figure*}[htbp]
        \centering
    \begin{minipage}[t]{\columnwidth}
        \centering
        \includegraphics[width=\textwidth]{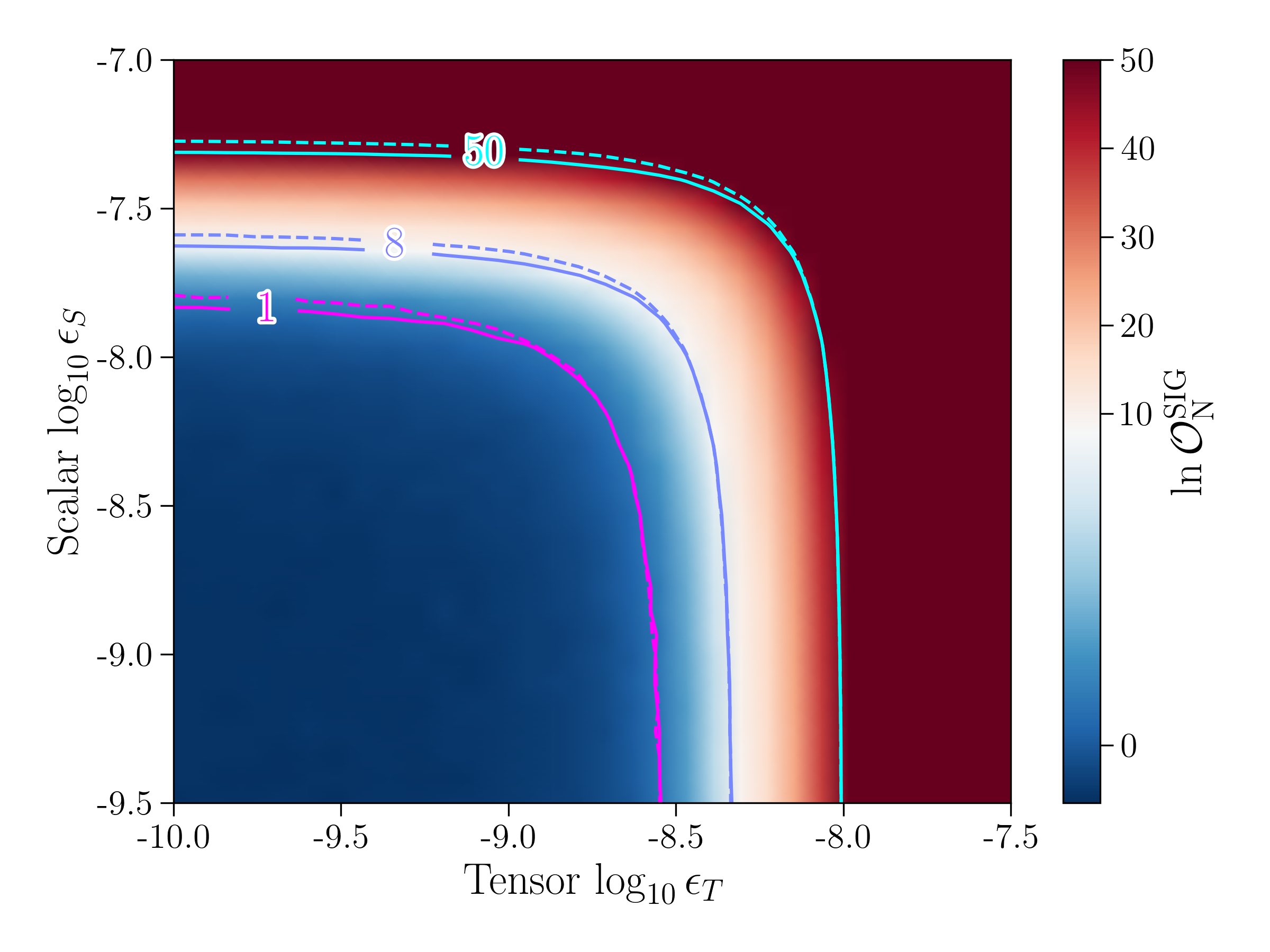}
    \end{minipage}
        \centering
    \begin{minipage}[t]{\columnwidth}
        \centering
        \includegraphics[width=\textwidth]{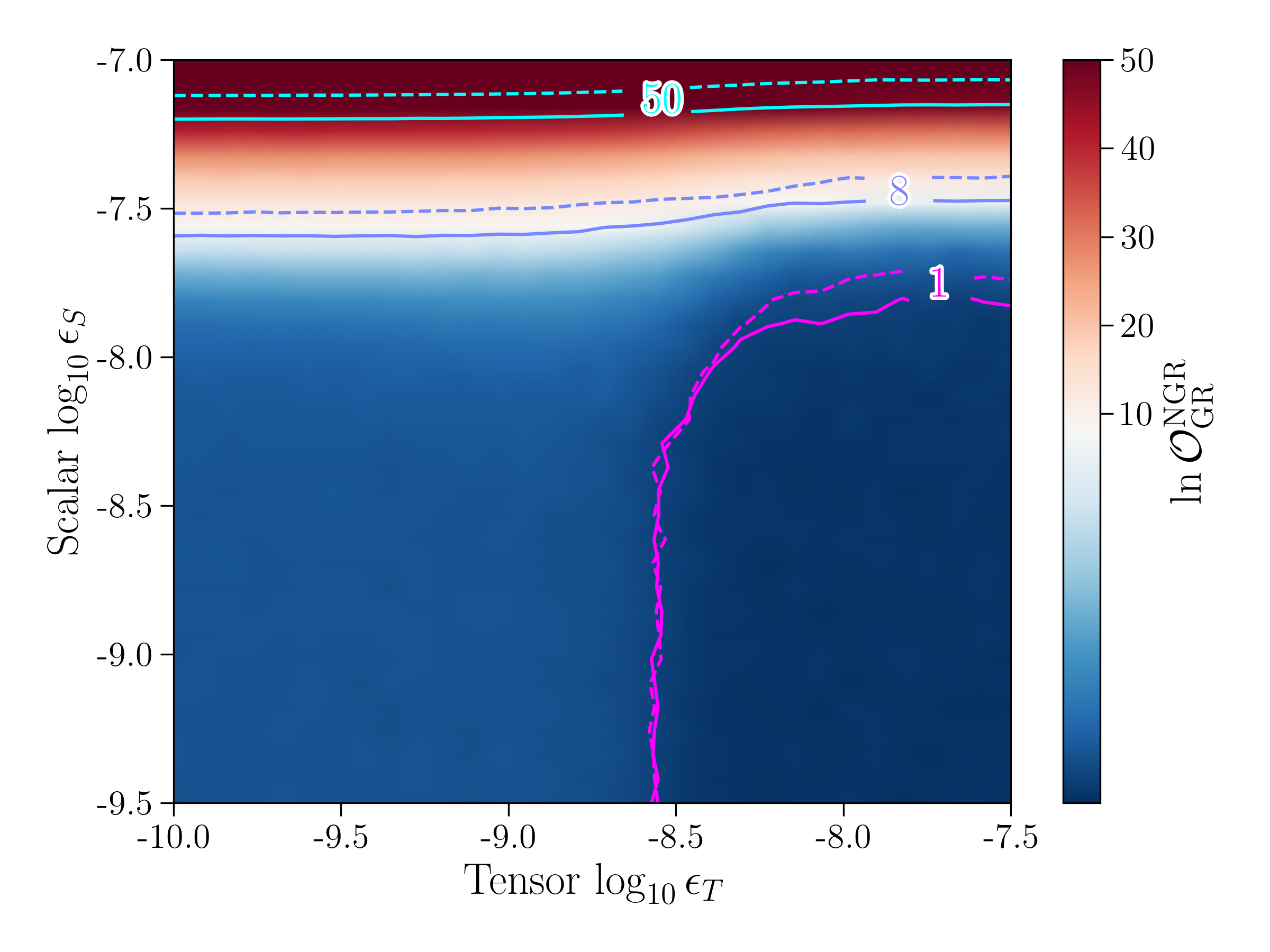}
    \end{minipage}
        \caption{\label{fig:mixpol}
        Left : two dimensional heat map of $\LnOSig$ between signal and
        noise hypotheses with the solid (dashed) contours of
        $\ln\mathcal{O}=1,8,50$ for \textit{Case 1} (\textit{Case 2}),
        respectively.        Right : two dimensional heat map of
        $\LnONgr$ between signal and noise hypotheses with the solid (dashed)
        contours of $\ln\mathcal{O}=1,8,50$ for \textit{Case 1} (\textit{Case
        2}), respectively.}
  \end{figure*}
To consider a more practical situation, we repeat the study described above
using an \ac{sgwb} injection with a mixture of the scalar and tensor
polarizations.  Following the configuration discussed in
\refref{tom_nongr_method}, for injected \ac{sgwb} signals, we adopt the
power-law $\bar{H}(f)$ spectrum with the power-law indices of $\alpha_T=2/3$ and
$\alpha_S=0$. Throughout this study, the Galactic-plane $\pbar_{\ell m}$ model
is consistently used with $\lmax=0(7)$ for the tensor (scalar) polarization,
simulating a signal of an anisotropic scalar background overlying on top of an
isotropic tensor background. We vary the amplitude factor for each polarization
component such that $\log_{10}\epsilon_S$ ($\log_{10}\epsilon_T$) ranges from
-9.5(-10.0) to -7.0(-7.5), and for a given grid point of the $\log_{10}\epsilon$
parameter space, a synthetic \ac{sgwb} signal is injected into a realization of
the noise data used in \secref{sec:sigid-singlepol}.  Also, injecting such a
composite signal model complicates its implementation in terms of the
precomputed parts, which is more discussed in \appref{app:injection}.

We analyze each dataset using the power-law $\bar{H}(f)$ + the Galactic-plane
$\pbar_{\ell m}$ model with every possible polarization combination to calculate
$\LnOSig$ and $\LnONgr$, adopting the same prior \ac{pdf} for $\epsilon$ and
$\alpha$ as in \secref{sec:sigid-singlepol}.  Eventually, a full set of runs
produces a two-dimensional map of each quantity across the $\log_{10}\epsilon$
parameter space. This whole process is iterated for the two configurations of
the recovery signal model as follows: For a given polarization combination,
\begin{itemize}
  \item \textit{Case 1} : $\lmax=0$ for the tensor polarization and $\lmax=7$ for the others,
  \item \textit{Case 2} : $\lmax=0$ for both polarizations.
\end{itemize}

\figureref{fig:mixpol} show a two-dimensional heat map of $\LnOSig$ (left) and
$\LnONgr$ (right) respectively for \textit{Case 1} with solid contours of
$\ln\mathcal{O}=1,8,50$. As a comparison in both plots, a result for
\textit{Case 2} is represented by the dashed contours, each of which corresponds
to the solid one with the same color.  One can see that a larger $\epsilon_S$
yields greater $\LnOSig$ or $\LnONgr$ and that, more importantly, the contours
of \textit{Case 2} (solid) all shift slightly below those of \textit{Case 1}
(dashed). This indicates that at a given ($\epsilon_T$, $\epsilon_S$) point,
$\LnOSig$ and $\LnONgr$ both increase by incorporating higher \ac{sph} modes of
a scalar background into the recovery signal model. Regarding the $\epsilon_T$
dependency, $\LnOSig$ monotonically increases with $\epsilon_T$, while $\LnONgr$
rather decreases especially in the lower $\epsilon_S$ regime as the tensor
polarization component in the injected signal starts to dominate over the scalar
one.  Although this overall structure of the two plots simply reproduces the
results shown in \refref{tom_nongr_method}, the key result here is the
difference between \textit{Case 1} and \textit{Case 2}, which is also consistent
with the behavior discussed in \figref{fig:lnO_hist}.

\section{\label{sec:pe} parameter estimation}
Apart from the signal detection evaluated by the odds ratio in the previous
sections, parameter estimation is one of the meaningful products obtained by the
Bayesian analysis, enriching the scientific implication.  In this section, we
study the behavior of posterior distributions with and without an anisotropic
\ac{sgwb} injection, using different configurations for the choice of the
$\lmax$ values in recovery (e.g. $\lmax$=0 or 7) as well as a detector network.
Note that in the case of $\lmax=0$ our formalism reduces to the Bayesian
analysis for an isotropic \ac{sgwb}, and hence this demonstrates a comparison
between isotropic and anisotropic analyses. Throughout this study, we use a
realization of the same dataset as \secref{sec:sigid-singlepol}, where an
\ac{sgwb} signal synthesized from one or mixed \ac{gw} polarizations may be
present. For signal recovery, following
\referencesref{tom_nongr_method,nongr_sgwb_lvk}, we adopt the most agnostic
model in terms of the \ac{gw} polarizations, i.e.  $\mathcal{H}_\mathrm{TVS}$.
Every polarization component of any signal model in this study involves the
Galactic-plane $\pbar_\mathrm{\ell m}$ (denoted as $\pbar^\mathrm{GP}_\mu$) and
a power-law $\hf{}{}$ distribution, and hence the injection and recovery signal
models both take the form of 
\begin{align}
    \sum_A\epsilon_A \left(\frac{f}{\SI{25}{\hertz}}\right)^{\alpha_A} \orf{\mu}{A} \pbar^\mathrm{GP}_\mu.
\end{align}
Here, $A$ indexes a \ac{gw} polarization and runs across a set of those involved
in a given signal model. Per \ac{gw} polarization component, we take
$\epsilon_A$ and $\alpha_A$ as free parameters to infer, which results in a
joint posterior \ac{pdf} concerning the six parameters in total. 
Throught this study, we adopt the same prior \ac{pdf} for $\epsilon_A$ and
$\alpha_A$ as in \secref{sec:sigid-singlepol}

\subsection{\label{sec:pe-noise} Gaussian-noise test}
\begin{figure*}[htbp]
    \includegraphics[width=\textwidth]{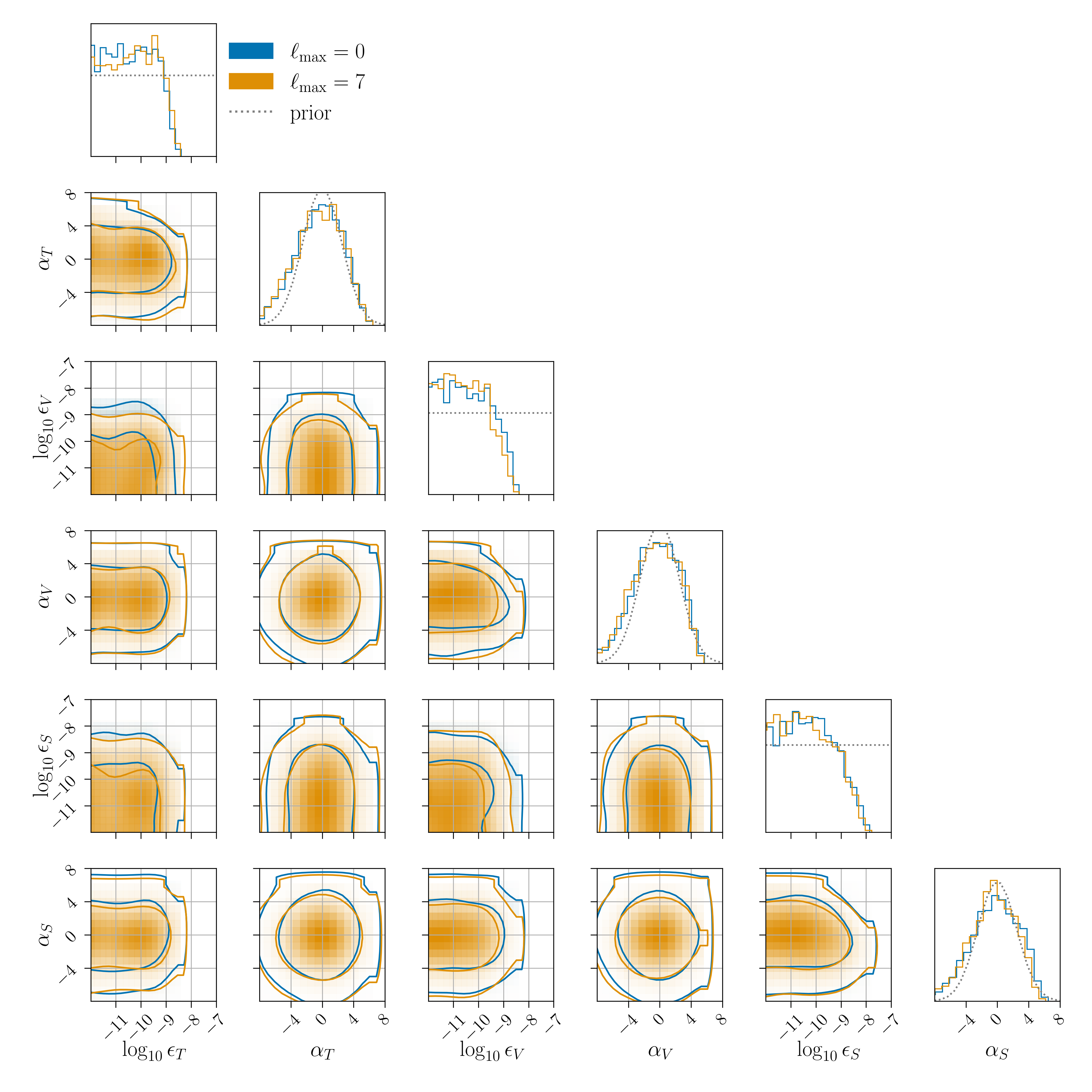}
    \caption{\label{fig:tvs_pe_noninj_HLV}
    Joint posterior \ac{pdf} of the Gaussian noise test using the HLV dataset
    analyzed with $\lmax=0$ (blue) and 7 (yellow) respectively, as well as a
    prior \ac{pdf} (gray dotted) mentioned earlier. The model parameters to
    infer in $\mathcal{H}_\mathrm{TVS}$ hypothesis consist of the amplitude
    factor $\epsilon$ and the power-law index $\alpha$ for each polarization
    (e.g.  tensor, vector, scalar from left to right in the $x$-axes)}
\end{figure*}

We first consider a situation where no \ac{sgwb} signal is present in a dataset of
the two \ac{ligo} detectors and Virgo (HLV). This analysis yields
$\LnOSig=-2.05(-2.14)$ and $\LnONgr=1.34(1.26)$ for $\lmax=0(7)$, suggesting a
nondetection of an \ac{sgwb} signal.  \figureref{fig:tvs_pe_noninj_HLV} shows a
joint posterior \ac{pdf} of the analysis with $\lmax=0$ (blue) and 7 (yellow)
respectively, as well as a prior \ac{pdf} (gray dotted) mentioned earlier.  One
can observe that the overall structure of the two posterior \acp{pdf} is largely
consistent. Also, across all the polarizations extremely larger values of
$\alpha$ are less likely than the smaller values because a frequency spectrum
with such a steep slope would have been detected if exists.  The null results of
each run place an upper limit on the amplitude factor, $\epsilon$, which is
summarized in \tableref{tab:epsilon_ul}. We note that although the upper limits
derived from the two analyses vary by around $\mathcal{O}(10)\%$, these values
should not be compared quantitatively as they are conceptually different
searches, targeting isotropy or Galactic-place structure.
%   \cleardoublepage

\subsection{\label{sec:pe-singlepol} Single polarization injection}
\begin{figure*}[htbp]
    \includegraphics[width=\textwidth]{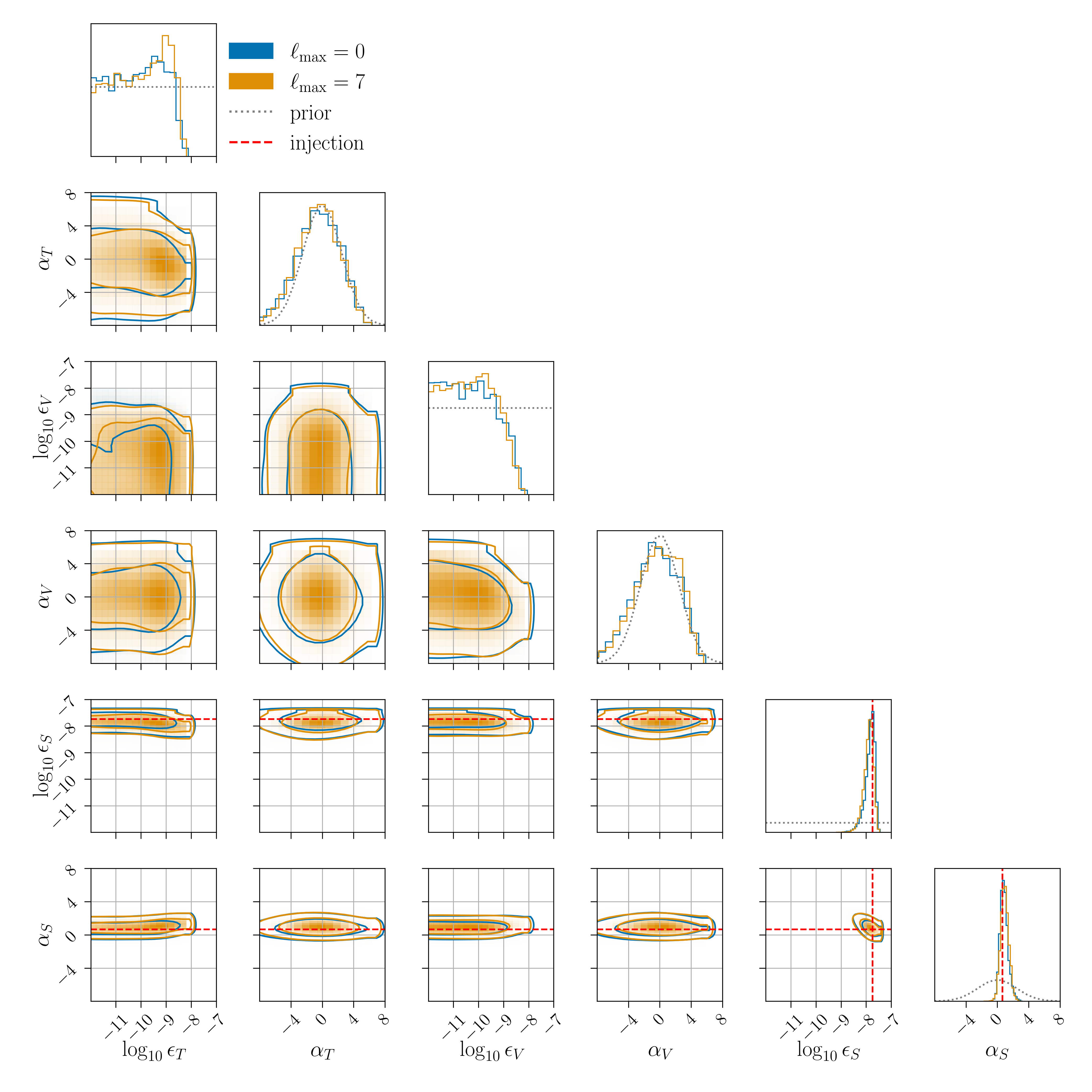}
    \caption{\label{fig:tvs_pe_scalarinj_HLV}
    Joint posterior \ac{pdf} of the scalar injection test using the HLV dataset
    analyzed with $\lmax=0$ (blue) and 7 (yellow) respectively, as well as a
    prior \ac{pdf} (gray dotted) mentioned earlier. The model parameters to
    infer in $\mathcal{H}_\mathrm{TVS}$ hypothesis consist of the amplitude
    factor $\epsilon$ and the power-law index $\alpha$ for each polarization
    (e.g.  tensor, vector, scalar from left to right in the $x$-axes)}
\end{figure*}

At this step, we simulate an anisotropic \ac{sgwb} signal with the scalar
polarization, adopting the power-law $\bar{H}(f)$ and the Galactic-plane
$\pbar_{\ell m}$ model with $\lmax=7$. Similar to the injected signals described
in \secref{sec:sigid-singlepol}, the power-law index $\alpha_S$ is set to 2/3
and the amplitude factor is adjusted so that the isotropic component of the
signal can be recovered with the optimal \ac{snr}=5 for the given dataset, e.g.
$\epsilon_S\approx\SI{1.8e-8}{}$. Analyzing the same HLV dataset as the
Gaussian-noise test in the presence of such an \ac{sgwb} injection, we obtain
$\LnOSig$ and $\LnONgr$  summarized in \tableref{tab:singlepol_odds} for each
$\lmax$ value. Both odds ratios indicate a great statistical significance and
even increase for the $\lmax=7$ case by around 3, which is consistent with
\figref{fig:lnO_hist}. This implies that an analysis using $\mathcal{H}_{TVS}$
hypothesis can properly obtain additional information from higher \ac{sph} modes
imprinted in the data.  \figureref{fig:tvs_pe_scalarinj_HLV} shows posterior
\acp{pdf} for each $\lmax=0,7$ case with the prior \ac{pdf} and the injected
value (red dashed) overlaid. Both analyses consistently recover the injected
$(\epsilon_S,\alpha_S)$ values and the overall structure of the two \acp{pdf}
are in great agreement.

\begin{figure*}[htbp]
    \includegraphics[width=\textwidth]{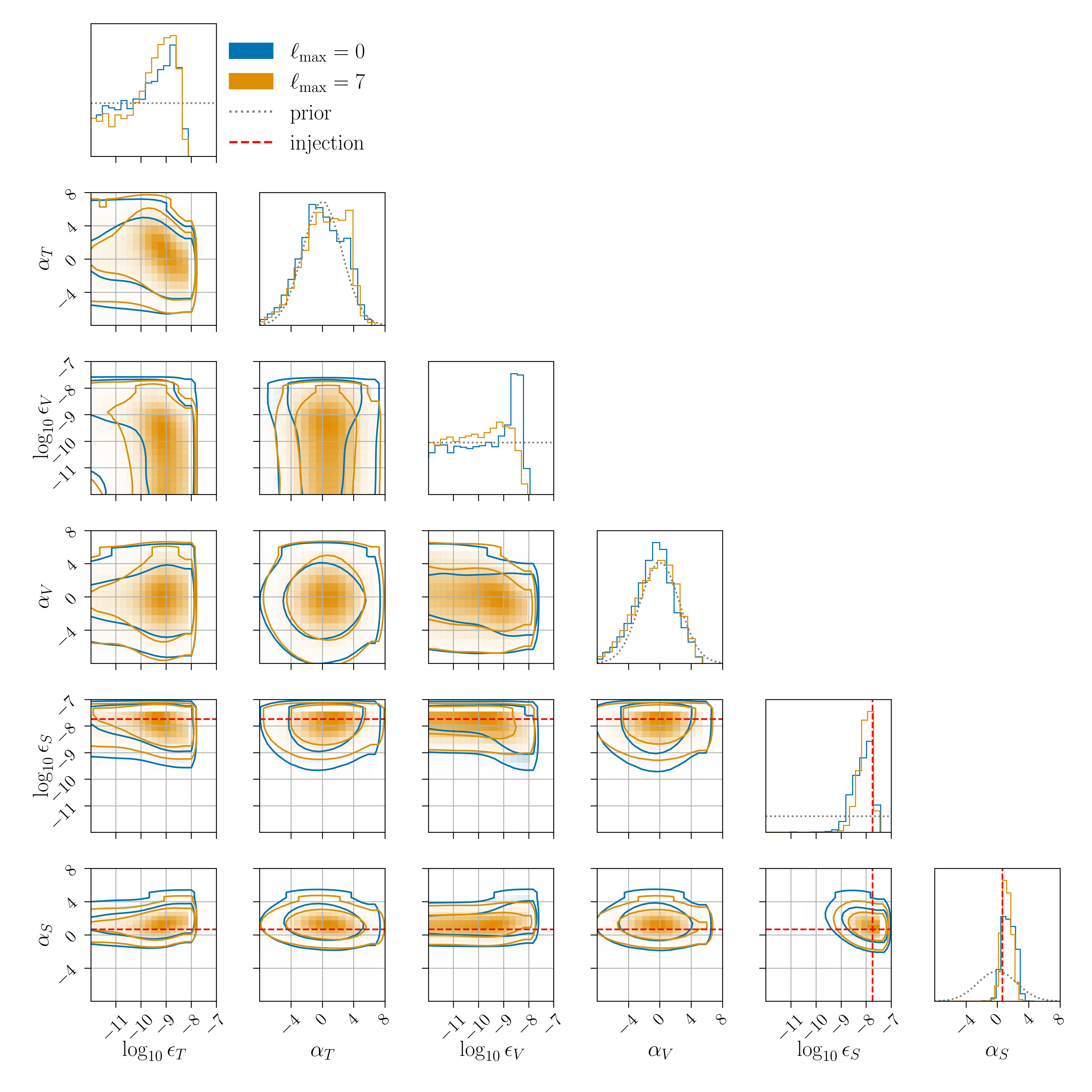}
    \caption{\label{fig:tvs_pe_scalarinj_HL}
    Joint posterior \ac{pdf} of the scalar injection test using the HL dataset
    analyzed with $\lmax=0$ (blue) and 7 (yellow) respectively, as well as a
    prior \ac{pdf} (gray dotted) mentioned earlier. The model parameters to
    infer in $\mathcal{H}_\mathrm{TVS}$ hypothesis consist of the amplitude
    factor $\epsilon$ and the power-law index $\alpha$ for each polarization
    (e.g.  tensor, vector, scalar from left to right in the $x$-axes)}
\end{figure*}

As a comparison, we repeat this study using a dataset of the two \ac{ligo}
detectors (HL) with the same duration and sensitivity curve, which results in
decreasing $\LnOSig,\LnONgr$ consistently for both $\lmax$ values (see
\tableref{tab:singlepol_odds}). Therefore, for the given dataset the addition of
Virgo helps the pipeline to better identify the signal and distinguish between
the NGR and GR hypotheses. Similar to \figref{fig:tvs_pe_scalarinj_HL}, the results
of posterior \acp{pdf} given by this HL dataset are shown in
\figref{fig:tvs_pe_scalarinj_HLV}. One of the noticeable features compared to
\figref{fig:tvs_pe_scalarinj_HL} is the poor precision of recovered
$(\epsilon_S, \alpha_S)$ \ac{pdf}, which further supports the gain from the
Virgo's data.
Also, there exist apparent peaks in the \acp{pdf} of $\epsilon_T$
and $\epsilon_V$, which reproduces the degeneracy between \ac{gw}
polarizations described in \refref{tom_nongr_method}.

Regarding the anisotropic component of the signal, one can evaluate its
detectability by comparing $\LnOSig$ between $\lmax=7$ and $0$, namely
\begin{align}
    \LnOAniso = \LnOSig|_{\ell=7} - \LnOSig|_{\ell=0}.
\end{align}
According to \tableref{tab:singlepol_odds}, $\LnOAniso=1.16$ and $0.67$ for HL and
HLV dataset respectively, which does not indicate strong evidence of the
anisotropic components given an expected \ac{snr} of the injected signal.
Nevertheless, it is worth noting that the $\lmax=7$ case mitigates the peak in
$\epsilon_V$'s \ac{pdf}. This can be understood by the fact that additional
\ac{sph} modes lead to characteristic \acp{orf} differing across \ac{gw}
polarizations in a signal model and provide extra consistency checks. In
consequence, the precision of $(\epsilon_S, \alpha_S)$ recovery slightly
improves compared to the $\lmax=0$ case.

\subsection{\label{sec:pe-mixpol} Mixed polarization injection}
\begin{figure*}[htbp]
    \includegraphics[width=\textwidth]{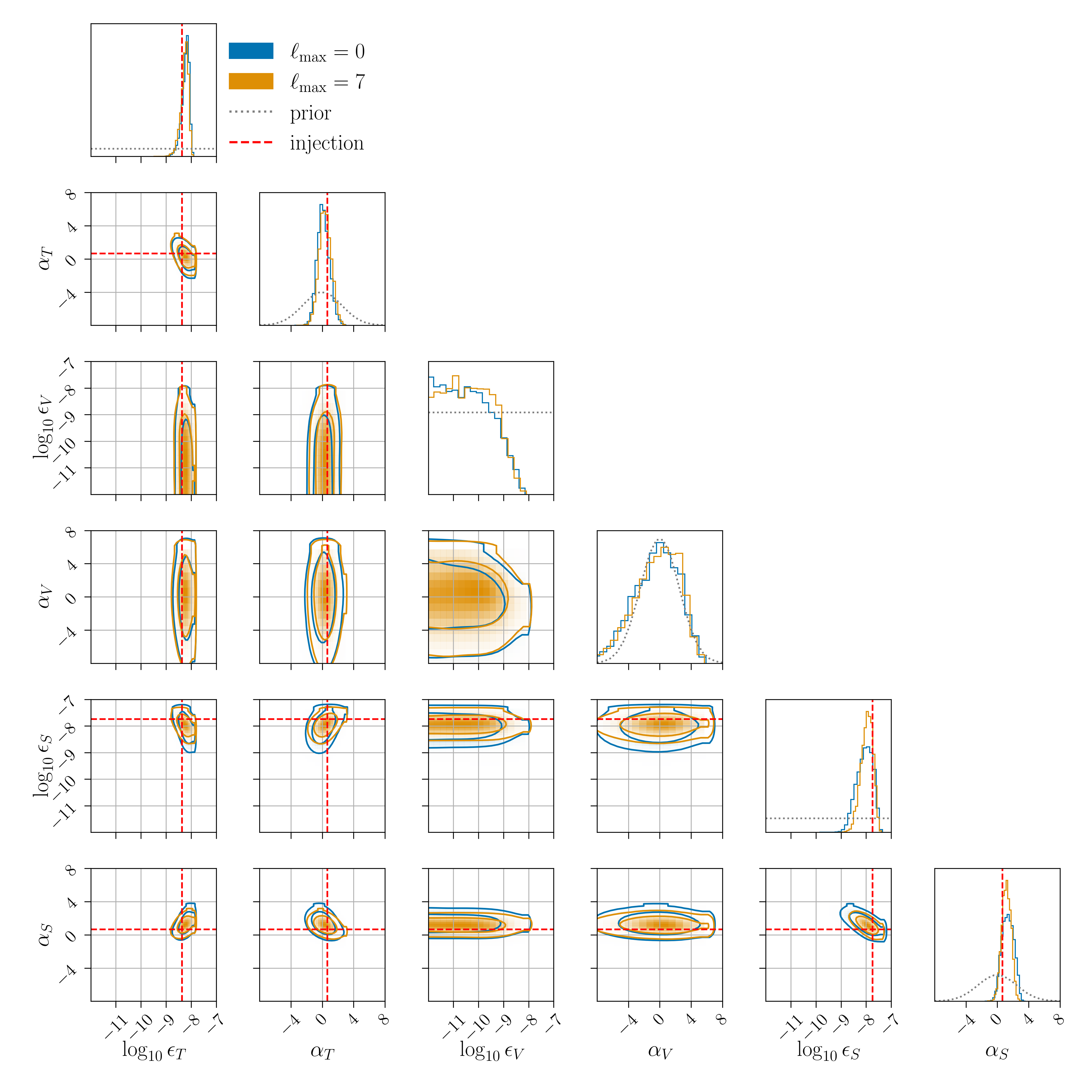}
    \caption{\label{fig:tvs_pe_tensorinj-scalarinj_HLV}
    Joint posterior \ac{pdf} of the scalar-tensor mixed injection test using the
    HLV dataset analyzed with $\lmax=0$ (blue) and 7 (yellow) respectively, as
    well as a prior \ac{pdf} (gray dotted) mentioned earlier. The model
    parameters to infer in $\mathcal{H}_\mathrm{TVS}$ hypothesis consist of the
    amplitude factor $\epsilon$ and the power-law index $\alpha$ for each
    polarization (e.g.  tensor, vector, scalar from left to right in the
    $x$-axes)}
\end{figure*}

Lastly, we further complicate a situation by mixing the scalar and tensor
polarizations in the injected signal model. Following the injection model in
\secref{sec:pe-singlepol}, we consistently use the power-law $\bar{H}(f)$ and
the Galactic-plane $\pbar_{\ell m}$ model with $\lmax=7$ for the \textit{both}
polarization components. Also, for each component, the power-law index is set to
2/3 and the amplitude factor is adjusted so that the isotropic component of the
signal can be recovered with the optimal \ac{snr}=5 for the given dataset, i.e.
$\epsilon_S\approx\SI{1.8e-8}{}$ and $\epsilon_T\approx\SI{4.3e-9}{}$.  With
this signal injected into the HLV dataset, \tableref{tab:mixpol_odds} shows that
$\LnOSig$ is significantly larger than those in \tableref{tab:singlepol_odds} for
both $\lmax=0$ and 7 cases due to an additional polarization component in the
injection, whereas $\LnONgr$ is largely consistent with
\tableref{tab:singlepol_odds}. Although, similar to \secref{sec:pe-singlepol},
$\LnOAniso$ derived from \tableref{tab:mixpol_odds} (3.4 and 3.5 for HL and HLV
dataset, respectively) does not show strong evidence of the anisotropic
components, the posterior \ac{pdf} shown in
\figref{fig:tvs_pe_tensorinj-scalarinj_HLV} indicates that a consistent recovery
of the four injected parameters for both cases and that, for the \acp{pdf} of
$\epsilon_S$ and $\alpha_S$, the $\lmax=7$ case produces slightly more precise
estimates than the $\lmax=0$ case.

\begin{figure*}[htbp]
    \includegraphics[width=\textwidth]{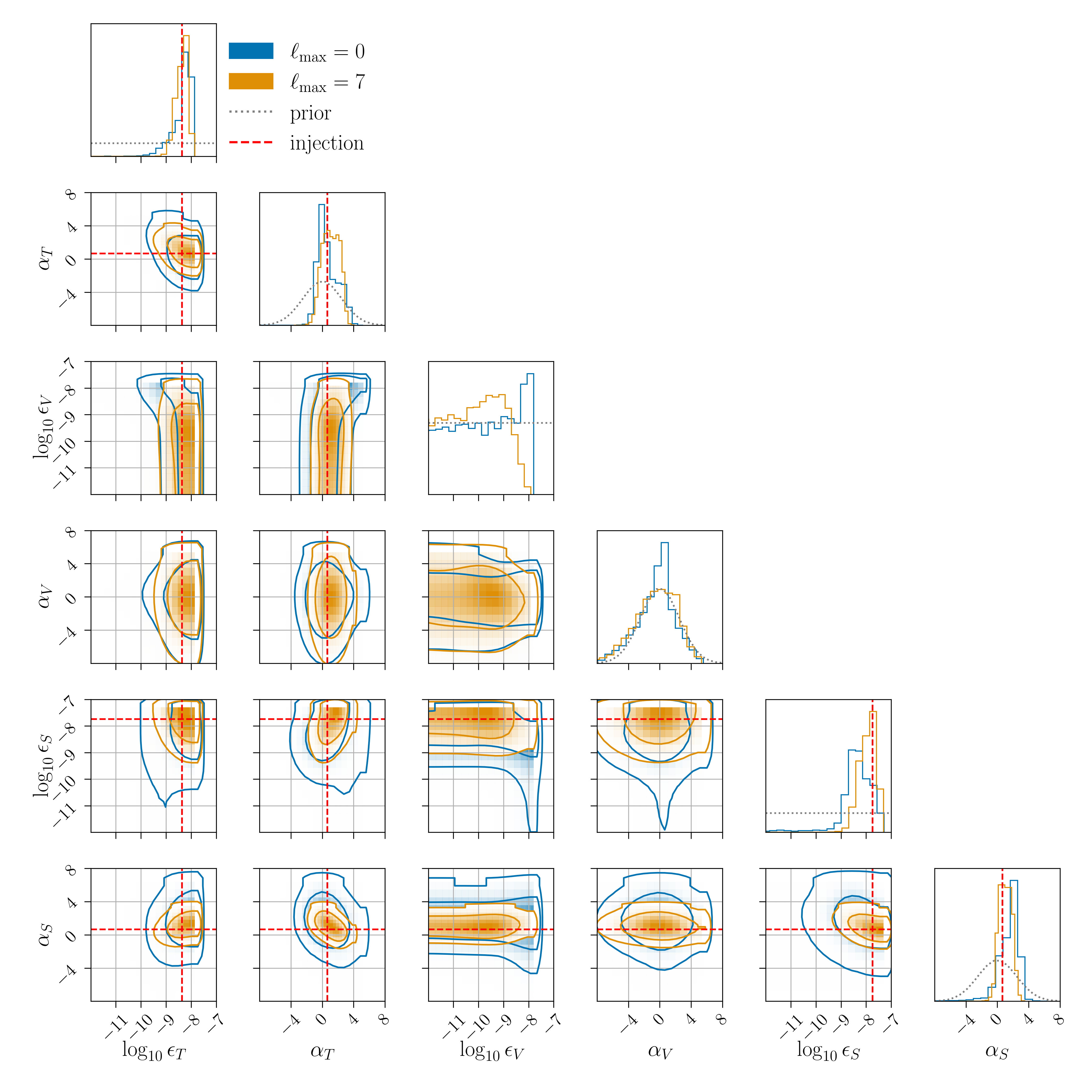}
    \caption{\label{fig:tvs_pe_tensorinj-scalarinj_HL}
    Joint posterior \ac{pdf} of the scalar-tensor mixed injection test using the
    HL dataset analyzed with $\lmax=0$ (blue) and 7 (yellow) respectively, as
    well as a prior \ac{pdf} (gray dotted) mentioned earlier. The model
    parameters to infer in $\mathcal{H}_\mathrm{TVS}$ hypothesis consist of the
    amplitude factor $\epsilon$ and the power-law index $\alpha$ for each
    polarization (e.g.  tensor, vector, scalar from left to right in the
    $x$-axes)
    }
\end{figure*}

The posterior \acp{pdf} derived from the HL dataset, as shown in \figref{fig:tvs_pe_tensorinj-scalarinj_HL}, exhibit a rather outstanding difference between these cases.
Specifically, the analysis with
$\lmax=7$ infers the injected parameter, $\epsilon_S$ and $\alpha_S$ in
particular, more precisely than the $\lmax=0$ case. Also, some of the parameters
among different polarization components present degeneracies for the $\lmax=0$
case, which is manifested as apparent peaks in $\epsilon_V$ and $\alpha_V$'s
\acp{pdf}, whereas the $\lmax=7$ case breaks the degeneracies and results in the
more precise inference of these parameters.  These findings, together with the
injection studies in \secref{sec:sigid-singlepol} and \secref{sec:pe-singlepol}, indicate
that in the presence of anisotropic \ac{sgwb} signal the inclusion of higher
\ac{sph} modes allows us not only to better identify signals but also to
estimate signal parameters more precisely without severe degeneracy between
them. Although in practice one does not know the right choice of $\lmax$ values
\textit{a priori}, \refref{sph_pe} demonstrates a systematic way to
optimize the choice of $\lmax$ values in terms of the odds ratio.

\begin{table*}[ht!] 
    \centering
{\tabcolsep = 0.3cm
{\renewcommand\arraystretch{1.55}
    \begin{tabular}{|c||c|c|c|}\hline
        & tensor ($\epsilon_T$) & vector ($\epsilon_V$) & scalar ($\epsilon_S$)\\ \hline\hline
      $\lmax=0$ & \SI{8.8e-10}{} & \SI{1.0e-9}{} & \SI{2.0e-9}{}  \\ \hline
      $\lmax=7$ & \SI{1.0e-9}{} & \SI{6.7e-10}{} & \SI{1.8e-9}{}  \\ \hline
    \end{tabular}
}}
    \caption{95\% upper limits on the amplitude factor $\epsilon$ for each \ac{gw} polarization obtained from the Gaussian noise test after marginalizing over the rest of the parameters.}
    \label{tab:epsilon_ul}
  \end{table*}
\begin{table*}[ht!]
    \centering
{\tabcolsep = 0.25cm
{\renewcommand\arraystretch{1.4}
    \begin{tabular}{|c||c|c|c|c|}\hline
      & \multicolumn{2}{c|}{$\lmax=0$} & \multicolumn{2}{c|}{$\lmax=7$} \\ \cline{2-5}
       & HL & HLV & HL & HLV\\ \hline\hline
      $\LnOSig$ & 12.49 & 15.35 & 13.65 & 16.02 \\ \hline
      $\LnONgr$ & 11.33 & 15.20 & 12.51 & 15.75 \\ \hline
    \end{tabular}
}}
    \caption{$\LnOSig$ and $\LnONgr$ obtained by the scalar injection test for each recovery $\lmax$ and detector network.}
    \label{tab:singlepol_odds}
\end{table*}
\clearpage
\begin{table*}[ht!]
    \centering
{\tabcolsep = 0.25cm
{\renewcommand\arraystretch{1.4}
    \begin{tabular}{|c||c|c|c|c|}\hline
      & \multicolumn{2}{c|}{$\lmax=0$} & \multicolumn{2}{c|}{$\lmax=7$} \\ \cline{2-5}
       & HL & HLV & HL & HLV\\ \hline\hline
      $\LnOSig$ & 42.22 & 43.95 & 45.62 & 47.45 \\ \hline
      $\LnONgr$ & 10.27 & 12.38 & 13.50 & 15.75 \\ \hline
    \end{tabular}
}}
    \caption{$\LnOSig$ and $\LnONgr$ obtained by the scalar-tensor mixed injection test for each recovery $\lmax$ and detector network.}
    \label{tab:mixpol_odds}
\end{table*}
\section{Conclusion}
In this work, we have described an extension of the Bayesian formalism searching
for an anisotropic \ac{sgwb} to incorporate nontensorial \ac{gw} polarizations.
In \secref{sec:orf-form}, we numerically compute the \ac{orf} for the scalar and
vector polarizations on the \ac{sph} basis, $\gamma_{\ell m}(f)$, which encodes
the sensitivity of a given detector pair to an anisotropic component of an
\ac{sgwb} characterized by the \ac{sph}'s indices $(\ell, m)$.  Adopting these
\acp{orf}, we generalize the likelihood function derived in
\refsref{sph_pe,tom_nongr_method} such that it allows for multiple components
with different polarizations, $\pbar_{\ell m}$ or $\bar{H}(f)$ in the recovery
signal model.

To demonstrate the capability to detect a nontensorial anisotropic \ac{sgwb},
we perform several simulation studies involving signal injections
with one or a mixture of the \ac{gw} polarizations. The results shown in
\figref{fig:lnO_hist} reproduce the findings discussed in
\refref{tom_nongr_method} and further suggest that, compared to the $\lmax=0$
case the right choice of the $\lmax$ value improves the detectability and the
odd ratio of the NGR hypothesis, $\LnONgr$. We also examine the results of
parameter estimation with regard to the $\mathcal{H}_{TVS}$ hypothesis, using
either HL or HLV dataset with and without an anisotropic \ac{sgwb} injection.
Similar to the previous injection study, we compare the posterior \acp{pdf}
between the two cases of $\lmax=7$ (the optimal value) and $\lmax=0$ (the
isotropic model). As a result, we find that, in the presence of an anisotropic
\ac{sgwb}, the addition of higher \ac{sph} modes into a recovery signal model
helps to break the degeneracy among different polarization components even
without Virgo's data and to infer model parameters more precisely.

In principle, the Bayesian formalism described in this work can be applied to
any signal model. An example of an astrophysically motivated $\pbar_{\ell m}$
distribution would be a population of millisecond pulsars in the Milky Way
Galaxy, e.g. the $\pbar_{\ell m}$ model developed in \cite{deepali_targeted}.
This formalism can evaluate any non-GR polarization component originating from
the population and, if it is sufficiently prominent, separate it from the GR
component, as illustrated in \secref{sec:pe-mixpol}.  Practically speaking,
however, one cannot adopt arbitrary large $\lmax$ value or a recovery signal
model with numerous free parameters as those would make matrix computation or
sampling process costly. Also, this formalism is intended to target a particular
anisotropic distribution model by fixing $\pbar_{\ell m}$ coefficients \textit{a
priori}, but ultimately we would be interested in directly inferring each of the
coefficients and producing $\pbar_{\ell m}$-independent results. At present, the
$\pbar_{\ell m}$ inference for physically meaningful $\lmax$ value is
computationally infeasible for the same reason.  One of the promising approaches
to these issues might be to make use of a GPU during the likelihood evaluation,
which may involve high-dimensional matrices. We leave this as a potential avenue for future improvement.

Alternatively, anisotropies subject to the cosmic variance, e.g. the large scale structure, are better suited to be modeled by not $\pbar_{\ell m}$ distribution but the angular power spectrum, which is given by
\begin{align}
    C_\ell = \frac{1}{2\ell+1}\sum_{m} |\mathcal{P}_{\ell m}|^2.
\end{align}
Since the likelihood function for $C_\ell$ is highly nontrivial, the formalism for a $C_\ell$-targeted search is not straightforward and has not been established yet, which remains to be future work to pursue.
Despite these caveats, the formalism presented in
this work has paved the way to establish a generalized Bayesian analysis for an
\ac{sgwb} including anisotropies and nontensorial polarizations.

% \cleardoublepage
\begin{acknowledgments}
    The author is grateful for computational resources provided by the LIGO
    Laboratory and supported by National Science Foundation Grants PHY-0757058
    and PHY-0823459.  This material is based upon work supported by NSF's LIGO
    Laboratory which is a major facility fully funded by the National Science
    Foundation.  LIGO was constructed by the California Institute of Technology
    and Massachusetts Institute of Technology with funding from the National
    Science Foundation (NSF) and operates under cooperative agreement
    PHY-1764464. The author is supported by the National Science Foundation
    through OAC-2103662 and PHY-2011865.
\end{acknowledgments}

\appendix

\section{ORF ANALYSIS}
\label{app:orf}
As discussed in \secref{sec:orf-form}, the \ac{sph}-based \ac{orf} encodes the
sensitivity of a given detector pair to an anisotropic component of an \ac{sgwb}
characterized by the \ac{sph}'s indices $(\ell, m)$. This appendix is dedicated
for further investigation and provides useful insight on its behavior across
different detector pairs and $(\ell, m)$ modes.

\begin{figure*}[htbp]
    \includegraphics[width=\textwidth]{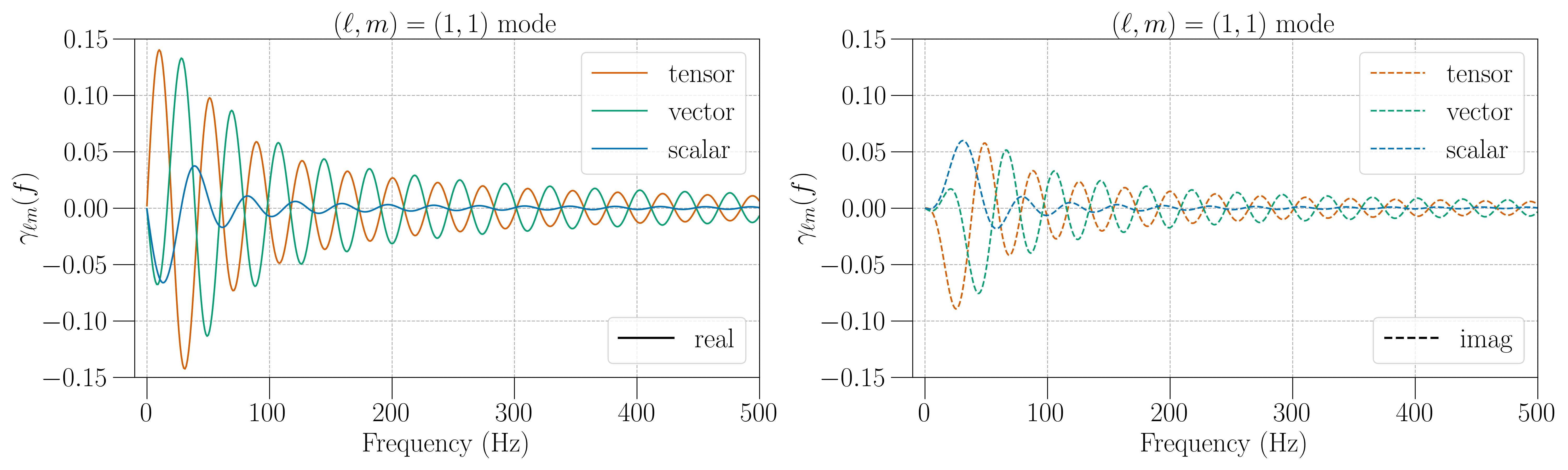}
    \caption{\label{fig:orf_11_hv}
    \ac{sph}-based \ac{orf} of the LIGO-Hanford and Virgo detector pair for each
    polarization family over \SIrange{0}{500}{\hertz} for $(\ell, m)=(1,1)$
    mode. The left plot shows the real part of the \ac{orf}, while the right one
    shows its imaginary part.}
\end{figure*}

\figuresref{fig:orf_11_hv}{fig:orf_22_hv} show real and imaginary parts of the
\ac{sph}-based \ac{orf} in $(\ell, m)=(1,1)$ and $(2,2)$ modes, respectively,
for the LIGO-Hanford and Virgo (HV) detector pair.  Similar to the monopole \ac{orf}
discussed in \refref{tom_nongr_method}, the intervals between zeros are much
shorter than those of HL detector pair shown in
\figsref{fig:orf_11_hl}{fig:orf_22_hl} because these intervals are
disproportional to a separation between the two detectors.
Also, over the whole frequency range, HV's \acp{orf} are smaller than HL's
by a factor of several, indicating that the geometry and orientation of the HV
pair makes it less sensitive to those \ac{sph} modes.

In contrast, the \ac{orf} for $(\ell, m)=(10, 10)$ mode plotted in
\figsref{fig:orf_1010_hl}{fig:orf_1010_hv} show some promise for the HV pair
with the following features.  First, one can see that both HL and HV's \acp{orf}
have the global peak of their amplitude located far away from \SI{0}{\hertz}.
This reflects the fact that a higher-frequency component of a given signal has a
better spatial resolution of its source. Second, the peak
location for HV's \ac{orf} is shifted slightly toward lower frequencies
($\sim\SI{100}{\hertz}$).  This can be explained by the diffraction-limit
argument discussed in the literature (e.g.
\cite{s5_directional,o1_directional,o2_directional,o3_directional}), where the
angular resolution on the sky is determined by the separation between detectors
($D$) and the most sensitive frequency ($f$):
\begin{align}
    \theta=\frac{c}{2Df}.
\end{align}
In other words, the typical frequencies sensitive to the given $\theta$ is
disproportional to the separation. Third, most importantly, the overall
amplitude of HV's \ac{orf} at around \SI{100}{\hertz} is larger than HL's,
particularly for the scalar and vector polarizations. Although the
diffraction-limit argument has qualitatively described this behavior, this
investigation more clearly demonstrates the advantage of the Virgo detector in
terms of the sensitivity to an anisotropic \ac{sgwb}. In reality, however, the
overall sensitivity of a given detector pair should account for the \ac{psd} of
each detector as well, which suppresses the benefit from HV's \ac{orf} given the
current projection of Virgo's sensitivity.

\begin{figure*}[htbp]
    \includegraphics[width=\textwidth]{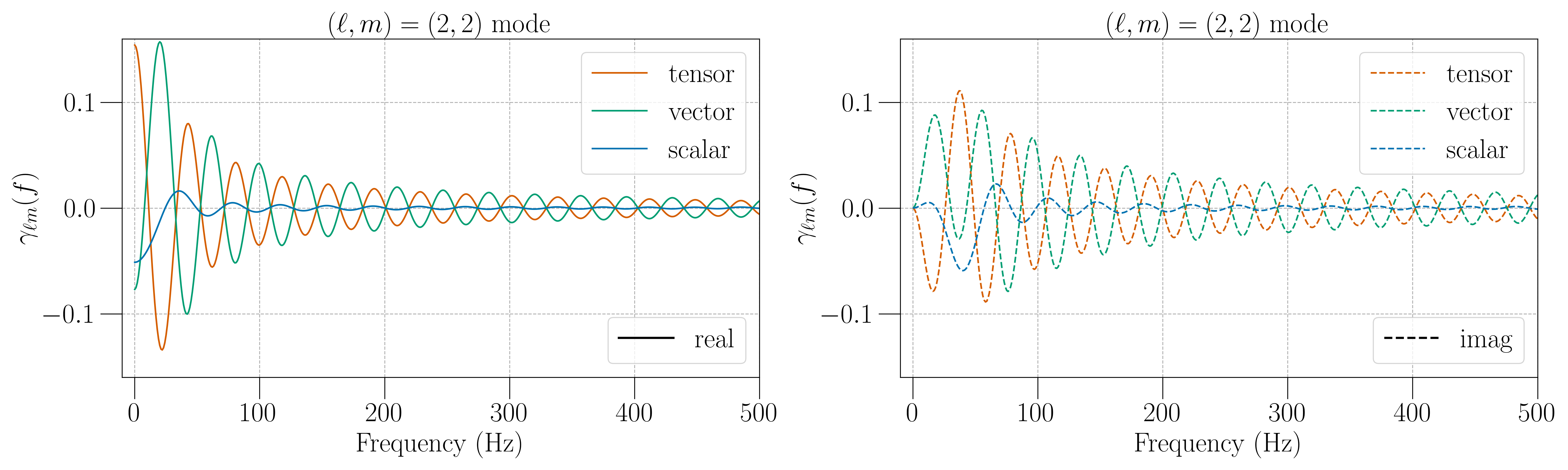}
    \caption{\label{fig:orf_22_hv}
    \ac{sph}-based \ac{orf} of the LIGO-Hanford and Virgo detector pair for each
    polarization family over \SIrange{0}{500}{\hertz} for $(\ell, m)=(2,2)$
    mode. The left plot shows the real part of the \ac{orf}, while the right one
    shows its imaginary part.}
\end{figure*}
\begin{figure*}[htbp]
    \includegraphics[width=\textwidth]{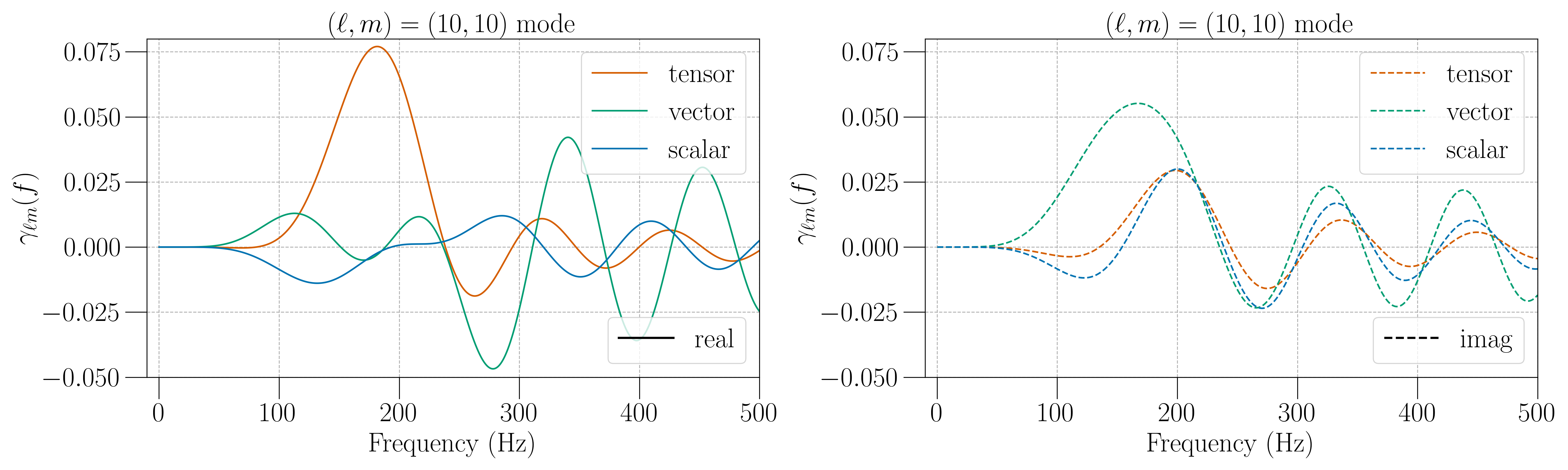}
    \caption{\label{fig:orf_1010_hl}
    \ac{sph}-based \ac{orf} of the two \ac{ligo}
    detector pair for each polarization family over \SIrange{0}{500}{\hertz} for
    $(\ell, m)=(10,10)$ mode. The left plot shows the real part of the \ac{orf},
    while the right one shows its imaginary part.}
\end{figure*}
\begin{figure*}[htbp]
    \includegraphics[width=\textwidth]{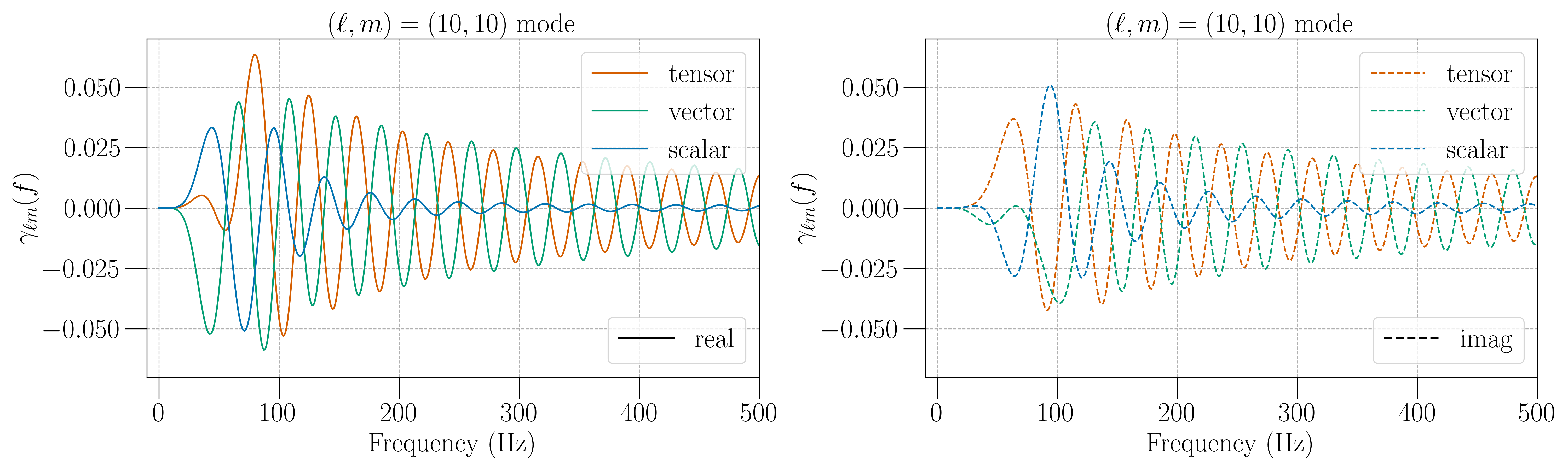}
    \caption{\label{fig:orf_1010_hv}
    \ac{sph}-based \ac{orf} of the LIGO-Hanford and Virgo detector pair for each
    polarization family over \SIrange{0}{500}{\hertz} for $(\ell, m)=(10,10)$
    mode. The left plot shows the real part of the \ac{orf}, while the right one
    shows its imaginary part.}
\end{figure*}
\section{INJECTING A COMPOSITE SIGNAL MODEL}
\label{app:injection} Here we describe our implementation of \ac{sgwb} injection
on the frequency and \ac{sph} domain in the presence of multiple components in a
signal model. To begin with, building upon Eq.~(23) in \refref{sph_pe},
injecting such a composite signal model into \ac{csd} reads
\begin{align}
  \label{eq:csd_inj}
  C(f, t)= C_n(f, t) + \sum_a\epsilon^\mathrm{inj}_a\hfinj{a}{\mathrm{inj}}\orf{\mu}{A_a}\pbar^{\mathrm{inj},a}_{\mu},
\end{align}
where $C_n(f, t)$ is \textit{noise-only} \ac{csd}, the superscript
$^\mathrm{inj}$ denotes each model or parameter intended for an injection and
the index runs across all the components in a given signal model for injection.
Substituting this expression into \eqref{eq:likelihood}, the expansion of its
exponent yields the following additional terms
  \begin{widetext}
  \begin{align}
    \label{eq:inj_terms}
    - \sum_a\epsilon^\mathrm{inj}_a\mathrm{Re}\left[(\pbar^{\mathrm{inj},a}_{\mu})^*X_\mu^{\mathrm{inj},a}\right]
    - \frac{1}{2}\sum_{a,b}\epsilon^\mathrm{inj}_a\epsilon^\mathrm{inj}_b(\pbar^{\mathrm{inj},a}_\mu)^*\fisher{\mu\nu}{\mathrm{inj},ab}\pbar^{\mathrm{inj},b}_\nu
    + \sum_{a,i}\epsilon^\mathrm{inj}_a\epsilon_i\mathrm{Re}\left[(\pbar^{\mathrm{inj},a}_\mu)^*\fisher{\mu\nu}{(c),ai}\pbar^{i}_\nu\right],
    % + \sum_{a,i}\epsilon\epsilon^\mathrm{inj}\pbar^*_\mu\fisher{\mu\nu}{(c)}\pbar_\nu^\mathrm{inj},
  \end{align}
  \end{widetext}
  due to the injection apart from the terms in \eqref{eq:rec_terms} arising from
  the recovery signal model. In \eqref{eq:inj_terms}, the indices $(a,b)$ run
  across all the components in a given signal model for injection, while the
  index $i$ denotes a component in a given signal model for recovery.
  $X_\mu^{\mathrm{inj},a}$ and $\fisher{\mu\nu}{\mathrm{inj},ab}$ are based on
  similar definitions to those shown in \eqsref{eq:dirty_map}{eq:fisher_matrix},
  respectively, replacing the indices $(i,j)$ with $(a,b)$ and $C$ with $C_n$.
  Also, $\fisher{\mu\nu}{(c),ai}$ is the \textit{coupled} Fisher matrix defined as 
    \begin{align}
      \label{eq:fisher_matrix_couple}
      \fisher{\mu\nu}{(c),ai} &=\sum_{f} \sum_{t} \orf{\mu}{A_a*} \frac{\tau\Delta f\bar{H}_i\bar{H}^\mathrm{inj}_a}{P_{1}(f, t) P_{2}(f, t)} \orf{\nu}{A_i},
    \end{align}
  generalizing Eq.~(25) in \refref{sph_pe}.
\normalem
\bibliography{references}% Produces the bibliography via BibTeX.

\end{document}